\documentclass[longauth]{aa}

\usepackage{mathrsfs}
\usepackage{url}
\usepackage{hyperref}
\hypersetup{
    colorlinks=true,
    linkcolor=blue,
    filecolor=black,      
    urlcolor=blue,
    citecolor=blue,
    pdftitle={Overleaf Example},
    pdfpagemode=FullScreen,
    }
\usepackage{comment}

\usepackage{graphicx}
\usepackage{txfonts}
\usepackage{lipsum}
\usepackage{subcaption}  
\usepackage{ulem}
\usepackage{xcolor}
                                
\usepackage{lscape}             
\usepackage{placeins}           
\usepackage{threeparttable}
\usepackage{threeparttablex}

\begin{document}

   \title{Unveiling the period-bounce population of cataclysmic variables}
   \subtitle{Spectroscopic and time-domain follow-up of eROSITA-selected candidates}

   \author{S. Hernández-Díaz\inst{1}
        \and B. Stelzer\inst{1} \and A. Schwope\inst{2} \and D. Muñoz-Giraldo\inst{1} \and M. R. Schreiber\inst{3} \and J. Brink\inst{2,4} \and K. G. Pradeep\inst{2,4} \and \newline B. T. Gänsicke\inst{5} \and M. Eracleous\inst{6} \and K. Szekerczes\inst{6} \and S. F. Anderson\inst{7} \and J. R. Brownstein\inst{8}
        }

\institute{Institut für Astronomie und Astrophysik, Eberhard Karls Universität Tübingen, Sand 1, 72076 Tübingen, Germany\\
\email{hernandez@astro.uni-tuebingen.de}
\and Leibniz Institut für Astrophysik Potsdam (AIP), An der Sternwarte 16, 14482 Potsdam, Germany
\and Universidad T\'ecnica Federico Santa Mar\'ia (UTFSM), Department of Physics, Av. España 1680, Valpara\'iso, Chile
\and Institute for Physics and Astronomy, University of Potsdam, Karl-Liebknecht-Str. 24/25, 14476 Potsdam, Germany 
\and Department of Physics, University of Warwick, Coventry, CV4 7AL, UK
\and Department of Astronomy and Astrophysics and Institute for Gravitation and the Cosmos, Penn State University, 525 Davey Lab, 251 Pollock Road, University Park, PA 16802, USA 
\and  Department of Astronomy, University of Washington, Box 351580, Seattle, WA 98195, USA 
\and Department of Physics and Astronomy, University of Utah, 115 S. 1400 E., Salt Lake City, UT 84112, USA
}
\date{Received September 30, 20XX}

 
\abstract
   {During their secular evolution, cataclysmic variable stars (CVs) evolve from long to short orbital periods ($P_{\rm orb}$) before reaching a minimum near $P_{\rm orb}\sim80$\,min. Beyond this stage, the donor departs from thermal equilibrium or becomes progressively degenerate, driving the system toward longer $P_{\rm orb}$. CVs that have evolved past this evolutionary turning point are known as period-bouncers (PBs). Despite predictions that 40–80\% of all CVs should be PBs, only 3–25\% of the observed CV population is actually composed of PBs. This discrepancy could possibly be attributed to their intrinsic low luminosities due to low mass-transfer rates.}
   {We aim to investigate the evolutionary status of 213 SRG/eROSITA-selected PB candidates through spectroscopic follow-up and time-domain observations. The sample also includes 19 previously confirmed PBs, which serve as benchmarks for evaluating the candidates.} 
   {We confirmed CVs among the PB candidates through the identification of Balmer emission lines in follow-up optical spectra from Sloan Digital Sky Survey-V (SDSS-V). Additionally, archival photometric data from multiple time-domain surveys, including \textit{Zwicky} Transient Facility (ZTF), Asteroid Terrestrial-impact Last Alert System (ATLAS), All-Sky Automated Survey for Supernovae (ASAS-SN), and Catalina Real-time Transient Survey (CRTS), were used to construct long-term light curves and search for dwarf-nova outbursts.

   By fitting hydrogen-rich atmosphere models to the SDSS-V spectra, we were able to estimate the effective temperature ($T_{\rm eff}$) and secular mass accretion rate ($\langle\dot{M}\rangle$) of the WDs. We also measured the Balmer decrements, used as diagnostics of the physical conditions of the accretion disc, to assess whether they are consistent with known PBs. In addition, we analysed archival two-minute cadence light curves from the Transiting Exoplanet Survey Satellite (TESS) to determine $P_{\rm orb}$ for a subset of systems, and compiled multi-wavelength photometry data to construct and model spectral energy distributions (SEDs), which we used to
infer approximate donor spectral types. Finally, we identified magnetic systems through Zeeman splitting in the Balmer lines and estimated their mean photospheric magnetic field strengths.}
   {We report the confirmation of 24 new CVs, comprising 18 systems identified through Balmer emission lines and 8 through outburst activity, with two systems independently identified by both methods. We constructed the SEDs for 14 of these systems, yielding donor $T_{\rm eff}<1850$\,K in 13 cases and $T_{\rm eff}\lesssim 2350$\,K for the remaining system. Our spectroscopic analysis of the new CVs observed with SDSS-V indicates that they are consistent with a PB nature, potentially increasing the population of confirmed PBs by $\sim 50\%$.
   }
   {Our results suggest that a substantial fraction of the PB population may remain hidden in WD catalogues. The combination of X-ray selection with spectroscopic and time-domain observations has led to the identification of additional systems that are very likely PBs. This strategy emerges as a powerful means to uncover these elusive systems, offering a systematic approach toward resolving the long-standing discrepancy between theoretical predictions and the observed PB population.}

   \maketitle

\nolinenumbers

\section{Introduction}
\label{Introduction}

Cataclysmic variables (CVs) are close binary systems in which a white dwarf (WD) accretes matter from a Roche-lobe–filling, low-mass main-sequence star, hereafter referred to as the donor (see \citealt{Warner_Book} for a comprehensive overview). The progenitors of CVs are binaries that evolve through a common-envelope phase. When the giant progenitor of the WD fills its Roche-lobe, dynamically unstable mass-transfer is triggered. The mass-transfer timescale becomes much shorter than the thermal timescale of the accretor, causing the transferred material to engulf both the core of the giant and its companion within a common envelope. The deposition of orbital energy and angular momentum into the envelope eventually drives its ejection, leaving behind a compact post-common-envelope system consisting of a WD and a low-mass main-sequence secondary (\citealt{Paczynski_1976}, \citealt{Zorotovic_2010}, \citealt{Ivanova_2013}). The onset of Roche-lobe overflow from the initially lower mass star may generate stable, angular momentum loss driven mass-transfer, which marks the formation of a CV (\citealt{Knigge_2011}, \citealt{Belloni_2022}).

The subsequent loss of angular momentum in the system, driven by magnetic braking (\citealt{Mestel_1968}, \citealt{Mestel_1987}, \citealt{Barraza_2025}, \citealt{Barraza_2026}) and gravitational wave radiation (GWR; \citealt{Faulkner_1971}, \citealt{Paczynski_1981}), gradually reduces the orbital separation, leading to shorter orbital periods ($P_{\rm orb}$) (\citealt{Rappaport_1982}, \citealt{Kolb_1993}, \citealt{Knigge_2011}, \citealt{Kalomeni_2016}). In systems below the period gap at $2-3$\,h (\citealt{Kolb_1998}, \citealt{Howell_2001}, \citealt{Schreiber_2024}), GWR is expected to dominate the angular momentum losses, driving the binary toward the period minimum at $P_{\rm orb}\sim80$\,min (\citealt{Rappaport_1982}, \citealt{Paczynski_1983}, \citealt{Knigge_2006}, \citealt{Knigge_2011}, \citealt{Patterson_2011}, \citealt{Goliasch_2015}). Beyond this point, the system evolves toward longer $P_{\rm orb}$ values, as the donor starts to expand under continued mass loss. This expansion occurs either because the donor is driven out of thermal equilibrium or because it becomes increasingly degenerate after its mass drops below the hydrogen-burning limit, leading to an inversion of its mass–radius relation \citep{Knigge_2006, Knigge_2011}. As the orbital separation increases, the donor also expands, which allows the system to sustain Roche-lobe overflow as $P_{\rm orb}$ evolves to larger values \citep{Warner_Book}. CVs in this evolutionary phase are referred to as period-bouncers (PBs; \citealt{Patterson_1998}, \citealt{Daniela_2024}).

As CVs evolve to and from the minimum period, they are expected to accumulate near it (\citealt{Gansicke_2009}) and contribute to the unresolved gravitational wave foreground of the Milky Way (e.g.,\citealt{Meliani_2000}, \citealt{Scaringi_2023}), detectable with Laser Interferometer Space Antenna (LISA; \citealt{Amaro_2017}). In fact, evolved CVs should impart very specific features to the spectral shape of the gravitational wave foreground (see Fig.~3 of \citealt{Scaringi_2023}).

Binary evolution models predict that PBs should dominate the CV population, comprising 40–80\% of systems (see e.g., \citealt{Kolb_1993}, \citealt{Goliasch_2015}, and \citealt{Belloni_2020}). However, this theoretical expectation is not supported by observations. To date, only 39 systems have been confirmed as PBs (\citealt{Daniela_2026}), and survey-based estimates indicate that PBs comprise about 3–25\% of the known CV population (\citealt{Inight_2023}, \citealt{Pala_2020}, \citealt{Rodriguez_2025}).

This discrepancy has motivated recent theoretical efforts aimed at reconciling predictions with observations. One possibility is that angular momentum loss during nova eruptions destabilises mass-transfer in CVs hosting low-mass WDs. As a result, many such systems may merge rather than evolve into PBs, reducing their expected numbers (\citealt{Schreiber_2016}, \citealt{Belloni_2018}). More recently, it has been suggested that CVs may become magnetic after reaching the period minimum (\citealt{Schreiber_2023}, \citealt{Schreiber_2025}). In this scenario, coupling between the magnetic fields of the WD and the donor leads to the transfer of spin angular momentum from the WD to the orbit, causing the system to detach.

Recently, \cite{Daniela_2026} proposed that a significant fraction of PBs may remain observationally hidden within WD catalogues because of their intrinsic faintness, a consequence of their low mass-transfer rates. Motivated by their association with X-ray sources detected by the extended ROentgen Survey with an Imaging Telescope Array \citep[eROSITA;][]{predehl2021} on board the Spektrum-Roentgen-Gamma mission \citep[SRG;][]{sunyaev2021}, \cite{Daniela_2026} identified 213 high-likelihood PB candidates within 500\,pc from the catalogue of \textit{Gaia} WD candidates by \citet{gentile2021}. These systems were selected on the basis of a multi-wavelength scorecard comprising ten observational parameters expected for PBs, together with X-ray emission consistent with bona fide PBs, that is, systems in which the donor has been detected spectroscopically or photometrically (see \citealt{Daniela_2026}). We note that this list of 213 high-likelihood PB candidates also contains 23 confirmed PBs.

Following the first confirmation of one of the eROSITA-selected high-likelihood PB candidates by \cite{Daniela_2026}, we present an analysis of follow-up optical spectroscopy from the Sloan Digital Sky Survey-V (SDSS-V) for a subset of the remaining unconfirmed candidates, with the aim of assessing their evolutionary status. The total sample studied includes 14 previously confirmed PBs which are part of the list of 213 eROSITA-selected high-likelihood PB candidates and serve as benchmarks for evaluating the remaining candidates. Additionally, we complement the SDSS-V spectroscopy with archival time-domain data and multi-wavelength photometry from various surveys, leading to the inclusion of five additional previously confirmed PBs in the study. Our approach follows the three observational criteria defined by \cite{Daniela_2026} for the confirmation of an eROSITA-selected candidate as a PB: (i) classification as a CV, (ii) determination of $P_{\rm orb}$ near the period minimum at $\sim$80 min, and (iii) constraints on the spectral type (SpT) of the donor consistent with a late-type, degenerate donor.

Beyond these core criteria, we conducted additional analyses to further evaluate the evolutionary state of our eROSITA-selected high-likelihood candidates. Based on SDSS-V spectroscopy, we modelled the WD component to derive effective temperatures and estimate secular mass accretion rates. We also measured the Balmer decrements as diagnostics of the physical conditions in the accretion discs of non-magnetic CVs, which differ systematically in PBs relative to pre-bounce CVs due to their lower mass accretion rates (see \citealt{Hernandez-Diaz_2026}). Finally, we identified four magnetic systems via Zeeman splitting and estimated their mean photospheric magnetic field strengths.

The sample of PB candidates and the observational database are described in Sect.~\ref{Candidates} and Sect.~\ref{Data base}, respectively. The analysis methodology is detailed in Sect.~\ref{Analysis}. In Sect.~\ref{Discussion}, we present and discuss our results, evaluating their implications for the PB status of our candidates. We summarise our findings and conclusions in Sect.~\ref{Summary}.

\section{eROSITA-selected period-bouncer candidates}
\label{Candidates}

Between December 2019 and December 2021, eROSITA completed four full-sky surveys (eRASS\,1–eRASS\,4). The corresponding source catalogues are generated by the eROSITA collaboration and distributed by the Max Planck Institute for Extraterrestrial Physics (MPE) using the eROSITA Science Analysis Software System \citep[eSASS;][]{brunner2022}. These catalogues include all detected eRASS sources in the western Galactic hemisphere (Galactic longitude $l \geq 180^\circ$), which corresponds to the sky area with German data rights. 

Our sample of 213 eROSITA-selected high-likelihood PB candidates within 500\,pc was produced by \cite{Daniela_2026} through a cross-match of the merged eRASS:4 catalogue (data processing version 020)—which combines data from the first four all-sky eRASS surveys—with \textit{Gaia}-selected WD candidates from \citet{gentile2021}. A reduced version of the multi-wavelength PB scorecard introduced by \citet{Daniela_2024}, together with specific X-ray selection cuts were then applied to identify systems with multi-wavelength properties and X-ray emission consistent with bona fide PBs. We obtained spectroscopic follow-up observations, as well as archival time-domain data (see Sect.~\ref{subsec:SDSS_Database}-\ref{subsec:multiwavelength_phot}) for these systems with the goal of closing in on their confirmation as PBs.

\section{Data base}
\label{Data base}

\subsection{SDSS-V optical spectroscopy}
\label{subsec:SDSS_Database}

The SDSS-V project is conducting a novel all-sky spectroscopic survey providing optical and infrared observations for millions of sources across the sky (\citealt{Kollmeier_2026}). Survey operations began in November 2020 with the 2.5\,m telescope at Apache Point Observatory (\citealt{Gunn_2006}) and have since expanded to include the 2.5\,m telescope at Las Campanas Observatory (\citealt{Bowen_1973}). Observations are conducted using the Baryon Oscillation Spectroscopic Survey spectrograph \citep[BOSS;][]{Smee_2013} and the Apache Point Observatory Galactic Evolution Experiment spectrographs \citep[APOGEE; ][]{Wilson_2019}. Each SDSS-V BOSS sub-exposure is obtained with an integration time of 900\,s, delivering a median resolution of $R\sim1800$ across the $3600-10000\,\AA$ wavelength range.

Out of the 213 eROSITA-selected high-likelihood PB candidates, 78 systems have been observed in SDSS-V up to MJD 61004. These observations were carried out within the Milky Way Mapper (MWM) program (\citealt{Kollmeier_2026}). Target selection for the MWM is organised into different cartons distinguished by their specific  selection function. Therefore, a given target can be included in different cartons; in our case, most of our targets were ingested into SDSS-V from the white dwarf carton (mwm\_wd\_pwd\_boss and mwm\_wd\_gaia\_boss) as well as from cartons that comprise CV candidates identified in the eROSITA eRASS1 and eRASS:3 catalogues. The three eROSITA CV cartons (mwm\_erosita\_compact\_gen, mwm\_erosita\_compact\_var, and mwm\_erosita\_compact\_boss) are described by \citet{Brink_2026}. Some of our targets overlap with those included in their work; however, they used an earlier cutoff date of MJD\,60708.

Access to these spectra enables the confirmation of new CVs (see Sect.~\ref{subsec:emission_lines}). In addition, they can subsequently be used to characterise the WDs (see Sect.~\ref{subsec:WD_model_fitting}) and the physical conditions of the accretion discs (see Sect.~\ref{subsec:Balmer_decrements}), thereby providing additional insights into the evolutionary status of the candidates (see Sect.~\ref{subsec:Teff_Mdot_Distributions} and Sect.~\ref{subsec:BD_discussion}).

\subsection{Long-term optical light curves}
\label{subsec:long-term_LCs}

We constructed long-term light curves by combining archival photometric data from multiple time-domain surveys. Specifically, we used data from the Catalina Real-time Transient Survey (CRTS; \citealt{drake2009}), the Asteroid Terrestrial-impact Last Alert System (ATLAS; \citealt{tonry2018}, \citealt{Heinze_2018}), the All-Sky Automated Survey for Supernovae (ASAS-SN; \citealt{shappee2014}, \citealt{Christy_2023}), and the \textit{Zwicky} Transient Facility (ZTF; \citealt{Masci_2019}). CRTS operated from 2007 to 2019, monitoring approximately 30,000 deg$^2$ of the sky ($-75^\circ < \delta < +65^\circ$), excluding the Galactic plane ($|b| \lesssim 10^\circ$–$15^\circ$), with each field observed in four exposures separated by $\sim$10 min and typically revisited in 2–4 observing sequences per lunation. ATLAS has operated since 2015 and, following the installation of its southern telescopes in 2021–2022, provides full-sky coverage with a cadence of $\sim$1 d between declinations $-50^\circ$ and $+50^\circ$ and $\sim$2 d in the polar regions, obtaining four 30\,s exposures over approximately one hour. ASAS-SN has operated since 2014, monitoring the entire observable sky ($\approx$30,000 deg$^2$) with a cadence of $\sim$2–3 d, which has improved to approximately one day since late 2017. ZTF has operated since 2018, surveying the northern sky visible from Palomar Observatory (down to declination $\approx30$) with a three-night cadence, while the Galactic plane ($|b|\leq7^\circ$, Dec. $\gtrsim -25^\circ$) is monitored nightly. Since 2020, the public northern-sky survey has adopted a two-night cadence.

Together, these surveys provide optical coverage spanning more than a decade for most of our 213 candidates, with only four systems having limited and highly sparse photometric data. This extensive coverage enables us to investigate long-term variability and identify outburst events (see Sect.~\ref{subsec:Outbursts}).

\subsection{TESS light curves}

The Transiting Exoplanet Survey Satellite (TESS; \citealt{TESS}) is a NASA mission led by the Massachusetts Institute of Technology (MIT) that conducts near all-sky photometric monitoring. The TESS observing strategy divides the sky into contiguous observing regions, referred to as sectors, each of which is monitored for $\sim$27 days, providing nearly continuous month-long light curves (\citealt{TESS}).

TESS has been proven to be well suited for the detection of periodicities in CV light curves (see e.g. \citealt{Bruch_5}, \citealt{Hernandez-Diaz_2025}, and \citealt{Meryem_2026}). In this work, we used TESS observations to search for orbital periods (see Sect.~\ref{subsec:Porb}). We cross-matched our sample of 213 eROSITA-selected high-likelihood PB candidates with the TESS Input Catalog (TIC; \citealt{TIC}) using J2000 equatorial coordinates within TOPCAT (\citealt{Topcat}) to obtain their corresponding TIC identifiers. We then retrieved two-minute cadence light curves from Sectors 1–98 from the Mikulski Archive for Space Telescopes (MAST; \citealt{mast_portal}). From our sample, 50 systems have TESS two-minute cadence light curves. 

\subsection{Multi-wavelength photometry}
\label{subsec:multiwavelength_phot}

We compiled multi-wavelength photometry spanning from the ultraviolet (UV) to the infrared (IR) to construct spectral energy distributions (SEDs; see Sect.~\ref{subsec:SEDs}), by cross-matching counterparts across different catalogues using an appropriate radius for each survey (see \citealt{Daniela_2024}). The aim of this analysis is to infer an approximate donor SpT from fitting atmosphere models to the SED. Constraining the donor mass (through its SpT) is a fundamental step in the confirmation of our candidates as PBs (see Sect.~\ref{subsec:SEDs} and Sect.~\ref{subsec:confirmation_PBs}).

The photometry was compiled from several large-scale surveys, including the Galaxy Evolution Explorer (GALEX; \citealt{Martin_2005, Morrissey_2007, Bianchi_2017}), \textit{Gaia} Data Release 3 (DR3; \citealt{Gaia_2016}, \citealt{Riello_2021}, \citealt{Gaia_2023}), SDSS (\citealt{Fukugita_1996}; \citealt{Smith_2002}), the VISTA Hemisphere Survey \citep[VHS;][]{Mahon_2013}, the UKIRT Infrared Deep Sky Survey \citep[UKIDSS;][]{Lawrence_2007}, the Two Micron All Sky Survey \citep[2MASS;][]{Skrutskie_2006}, and the Wide-field Infrared Survey Explorer (WISE) AllWISE data release (\citealt{Erward_2010}, \citealt{Mainzer_2011}). 

\section{Analysis}
\label{Analysis}

The aim of this study is to identify PBs among the eROSITA-detected candidates selected by \citet{Daniela_2026} from the \textit{Gaia} WD catalogue of \citet{gentile2021}. To this end, we performed a comprehensive analysis of the data base described in Sect.~\ref{Data base}. In the first step, we used the optical spectra and long-term light curves to find signatures that establish the object as a CV (see Sect.~\ref{subsec:Confirmation_as_CVs}). The CV nature is the first (and trivial) criterion that a PB must fulfill. The second criterion is the detection of $P_{\rm orb}$ near the period minimum at $\sim$80 min, which is investigated using TESS light curves (see Sect.~\ref{subsec:Porb}). To obtain evidence to support a late-type, degenerate donor as the third and final criterion for PB confirmation, we relied on modelling the SED (see Sect.~\ref{subsec:SEDs}) since the very cool companion does not produce a measurable signal in the optical spectra.

We complemented these three confirmation criteria with additional analyses to further assess the PB status of the systems. This included modelling the WD photospheric emission in the SDSS-V spectra to derive their effective temperatures and infer the corresponding secular mass accretion rates from compressional heating (see Sect.~\ref{subsec:WD_model_fitting}), together with obtaining measurements of the Balmer decrements to probe the physical conditions in the accretion discs of non-magnetic CVs (see Sect.~\ref{subsec:Balmer_decrements}). The Balmer decrements have recently been shown to distinguish PBs from pre-bounce CVs, reflecting the lower mass accretion rates characteristic of PBs (see \citealt{Hernandez-Diaz_2026}). We note that although these additional diagnostics provide independent evidence supporting the PB classification, they were not used as formal confirmation criteria. Finally, we identified Zeeman splitting in the SDSS-V spectra of four magnetic CVs, allowing us to measure their mean photospheric magnetic field strengths (see Sect.~\ref{subsec:Zeeman_Splitting}).

\subsection{Spectroscopic and photometric confirmation of CVs}
\label{subsec:Confirmation_as_CVs}

\subsubsection{Emission lines as signatures of ongoing accretion}
\label{subsec:emission_lines}

In non-magnetic CVs, the optical spectrum generally exhibits a blue continuum together with emission lines, primarily from the Balmer and Paschen series of hydrogen, as well as He I, He II, Ca II, and Fe transitions (e.g. \citealt{Smith_2006}). Most of this emission is produced in the accretion disc surrounding the WD. The emission lines are typically broad, corresponding to velocities of thousands of kilometres per second, and in systems with moderate to high inclination ($i \gtrsim 15^{\circ}$) they often exhibit double-peaked profiles, as they trace the Keplerian velocity field of the rotating gas (\citealt{Smak_1969}, \citealt{Smak_1982}, \citealt{Horne_1986}). Additional components of the system can contribute to the observed emission lines, including the gas stream transferring material from the donor (e.g. \citealt{Harlaftis_1996}), the impact region where the stream collides with the disc (the so-called hot spot; e.g. \citealt{Neustroev_2016}), and the irradiated hemisphere of the donor (e.g. \citealt{Stephen_1992}).

In magnetic CVs (mCVs), the formation of an accretion disc may be suppressed or absent, as the accretion flow is controlled by the magnetic field of the WD. In these systems, line emission can also originate in the magnetically channelled accretion columns (see \citealt{Cropper_1990} and \citealt{Patterson_1994} for a comprehensive review on mCVs).

Consequently, the detection of broad emission lines, with $\mathrm{H}\alpha$ typically the most prominent, is a strong indicator of ongoing accretion in the system and therefore provides evidence to support the identification of the source as a CV. 

\begin{figure}
    \centering
    \includegraphics[width=0.46\textwidth]{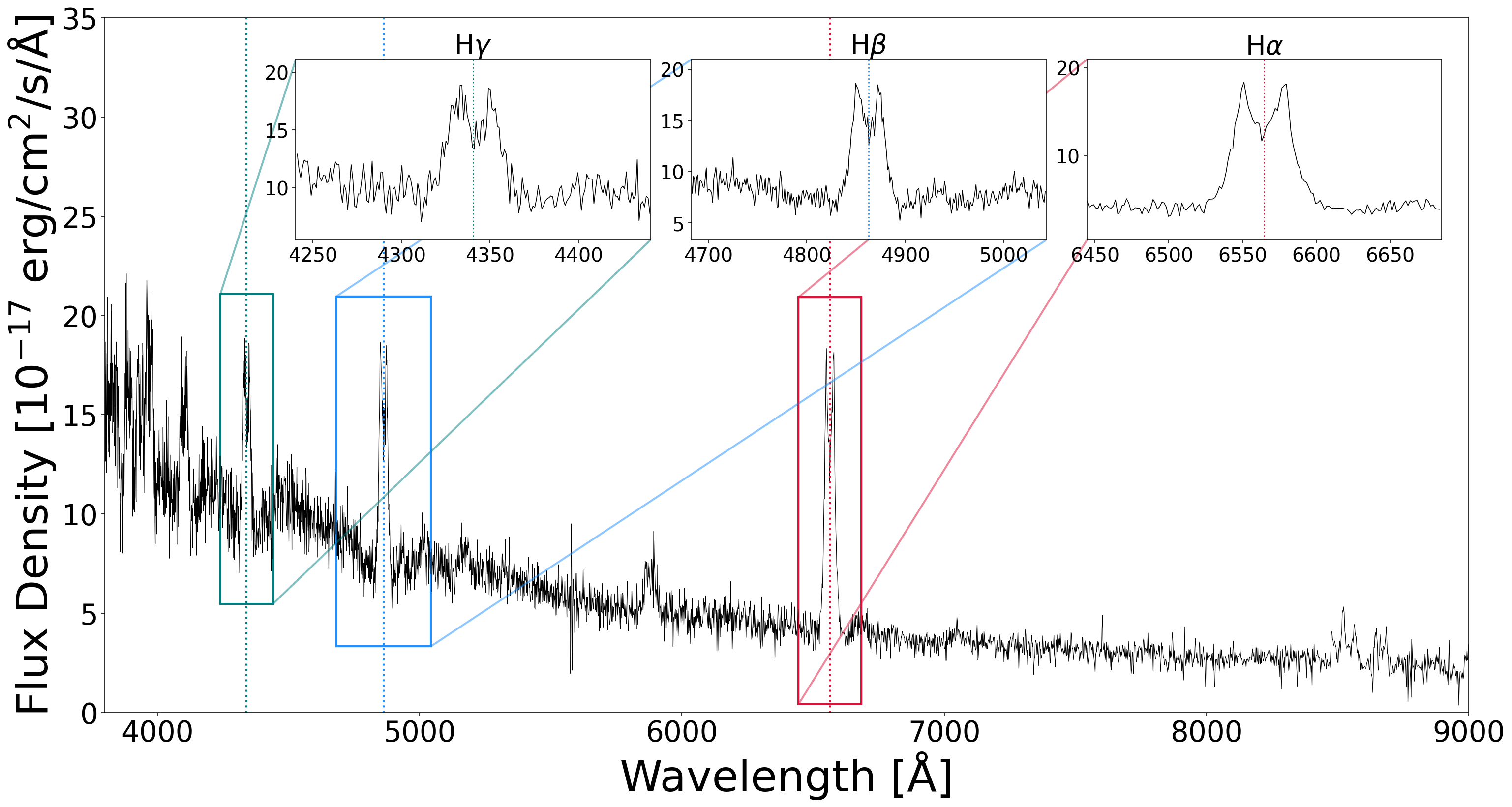}
    \vspace{0.6cm} 
    \includegraphics[width=0.46\textwidth]{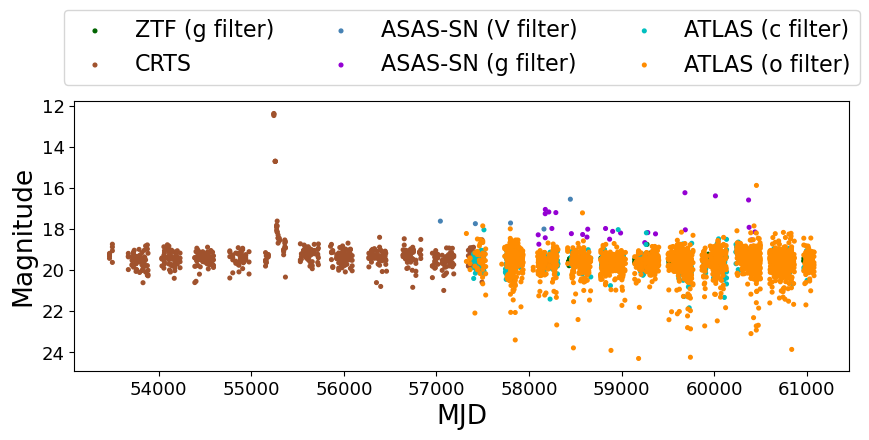}
    \caption{CRTS\,J104411.4+211307, a previously confirmed PB (\citealt{Daniela_2024}). Top: SDSS-V spectrum obtained at MJD 60704, displaying double-peaked Balmer emission lines. Bottom: Long-term light curves showing a superoutburst with an amplitude of $\sim$8\,mag at around MJD 55300.}
    \label{fig:Gaia DR3 3987650806740051712}
\end{figure}

We visually inspected the optical spectra of the 78 eROSITA-selected high-likelihood PB candidates observed in SDSS-V, identifying Balmer emission in 32 systems. Fourteen are previously confirmed PBs (see \citealt{Daniela_2026}). Fig.~\ref{fig:Gaia DR3 3987650806740051712} shows the SDSS-V spectrum of one of them as an example of a typical optical spectrum of a PB. The remaining $18$ systems with Balmer emission lines are newly identified CVs, the evolutionary status of which still needs to be understood through our subsequent analysis steps. A flag indicating whether a system has been observed in SDSS-V and exhibits Balmer emission is provided in Table~\ref{table:Master} (see Appendix~\ref{sec:master_table}). 
We find that all but one of the $32$ systems (with the exception being  GALEX\,J035124.8-092742) are also part of the cartons defined by \citet{Brink_2026} for SDSS-V follow-up.

\subsubsection{Dwarf-nova outbursts}
\label{subsec:Outbursts}

We examined the long-term light curves of the 213 eROSITA-selected high-likelihood PB candidates (see Sect.~\ref{subsec:long-term_LCs}) to identify dwarf-nova outbursts. Dwarf novae are non-magnetic CVs characterised by recurrent outbursts with amplitudes of $\sim$2–6 mag, durations of days to weeks, and recurrence times ranging from weeks to decades. These events arise from enhanced accretion onto the WD driven by thermal–viscous instabilities in the accretion disc. In short-period CVs below the 2–3\,h period gap, tidal instabilities or mass-transfer variations can also occur, producing superoutbursts of $\sim$8 mag (see \citealt{Osaki_1996}, \citealt{Lasota_2001} and \citealt{Schreiber_2004} for a comprehensive review).

The detection of such outbursts provides strong evidence to support a CV nature, as they directly trace accretion through a disc and therefore they distinguish the system from an isolated WD. In the context of PB candidates, their low mass-transfer rates result in long quiescent intervals (typically $\gtrsim$ decades) and high-amplitude outbursts. CVs displaying this variability behaviour are generally classified as WZ\,Sge-type dwarf novae, which constitute typical PB candidates (see \citealt{Kato_2015}). 

We found 24 eROSITA-selected high-likelihood PB candidates with at least one outburst in their long-term light curves. Sixteen of these were previously known as CVs, including five also observed in SDSS-V. Of these 16 systems, 10 were previously confirmed PBs (see \citealt{Daniela_2026}). For the remaining eight systems, the identification of outburst activity suggests a classification as a CV. For two of these, GALEX\,J040338.2-104945 and \textit{Gaia} DR3 2902110913736006144, the CV nature was also established through the SDSS-V spectra. An example of the typical dwarf-nova outburst observed in a PB  is shown in Fig.~\ref{fig:Gaia DR3 3987650806740051712}. In Table~\ref{table:Master} (see Appendix~\ref{sec:master_table}), a flag indicates whether a system was identified to display an outburst.

\subsubsection{AM\,CVns}
\label{subsec:AM_CVns}

As anticipated by \citet{Daniela_2026}, a fraction of our eROSITA-selected high-likelihood PB candidates are expected to be AM\,CVn systems. AM\,CVns are ultra-compact interacting binaries with orbital periods between 5\,min and 65\,min (\citealt{Ramsay_2018}) and helium dominated spectra characterised by the absence of detectable hydrogen lines. They consist of a WD accretor and a helium-rich donor, with mass-transfer driven by angular momentum losses through GWR (see \citealt{Nelemans_2005} for a comprehensive review). Given their different evolutionary channel, we do not analyse these systems in this work. However, we report those identified among our sample in Table~\ref{table:AM_CVns} (see Appendix~\ref{table:AM_CVns}).

Among the eROSITA-selected high-likelihood PB candidates observed in SDSS-V, we identified six AM\,CVn systems, four of which were not previously known. Five of these systems are also part of the SDSS-V follow-up cartons by \citet{Brink_2026}, and the two among them which were observed before their cutoff date (see Sect.~\ref{subsec:SDSS_Database}) are included in their catalogue. We note that the spectrum of GALEX\,J112611.6+154926 is noisy, and therefore its classification is tentative.

Of the 17 previously known CVs with an outburst in their long-term light curve (see Sect.~\ref{subsec:Outbursts}), two systems have been previously confirmed as an AM\,CVn. Finally, for \textit{Gaia} DR3 3124112584945910912, its long-term light curve reveals a superoutburst with an amplitude of $\Delta m \sim 8$. An Astronomer’s Telegram (\citealt{Maehara_2023}) reports a helium-dominated spectrum, suggesting that this system is also an AM\,CVn. If confirmed, it would rank among one of the most extreme AM\,CVn outbursts, whose amplitudes typically lie in the range $\sim$3.5–6 mag (\citealt{Kotko_2012}). The system was recently included in the catalogue of ultracompact WD binaries by \citealt{Green_2025}. We note that for systems showing outbursts but lacking spectroscopy, determining $P_{\rm orb}$ is essential to distinguish between hydrogen-rich CVs and AM\,CVns, as the latter have $P_{\rm orb}$ values below the hydrogen-rich CV period minimum at $\sim 80$\,min. Therefore, some of the newly identified CVs based on outbursts in their long-term light curves (see Sect.~\ref{subsec:Outbursts}) may correspond to AM\,CVn systems.

\subsection{Search for orbital periods}
\label{subsec:Porb}

In this work, we adopted the strategy of \cite{Hernandez-Diaz_2025} to determine $P_{\rm orb}$ using TESS two-minute cadence light curves. The methodology combines four period-search techniques—Lomb–Scargle periodogram, autocorrelation function (ACF), sine fitting, and Fourier power spectrum analysis—enabling cross-validation among different methods and thus improving the robustness of the detected signals.

For systems whose periods have been independently recovered in multiple TESS sectors, we determined (for each period-search method) the mean period across sectors and adopt the associated standard error as its uncertainty. The final adopted period is the one provided by the method yielding the smallest relative uncertainty. When a signal was detected in only one TESS sector, the period was adopted from the Lomb–Scargle periodogram, and its uncertainty was estimated through Monte Carlo resampling of the light curve (see \citealt{Hernandez-Diaz_2025}). 

To verify whether the TESS observations were obtained during quiescence, we compared their temporal coverage with the long-term light curves. As our aim is to derive $P_{\rm orb}$, we restricted the analysis to light curves obtained entirely during quiescence, resulting in period measurements for eight systems. The reliability of each detection is assessed using the S/N$_{\rm PSD}$ parameter introduced by \cite{Hernandez-Diaz_2025}, defined as the signal-to-noise ratio (S/N) of a frequency peak in the power spectral density (PSD) of a light curve. In that earlier work, we had determined a minimum S/N$_{\text{PSD}}$ threshold of $\left(\text{S/N}_{\text{PSD}}\right)_{\text{min}}=0.004$, below which period detections in TESS two-minute cadence light curves are considered unreliable. The final adopted $P_{\rm orb}$ values and the corresponding S/N$_{\rm PSD}$ measurements are reported in Table~\ref{table:orbital_periods} in Appendix~\ref{sec:orbital_periods_results}, with a flag indicating reliable detections.
We discuss the results from the analysis of TESS light curves in Sect.~\ref{subsec:periods}.

\subsection{Characterisation of the white dwarfs}
\label{subsec:WD_model_fitting}

\begin{figure*}
    \centering
    \includegraphics[width=0.9\linewidth]{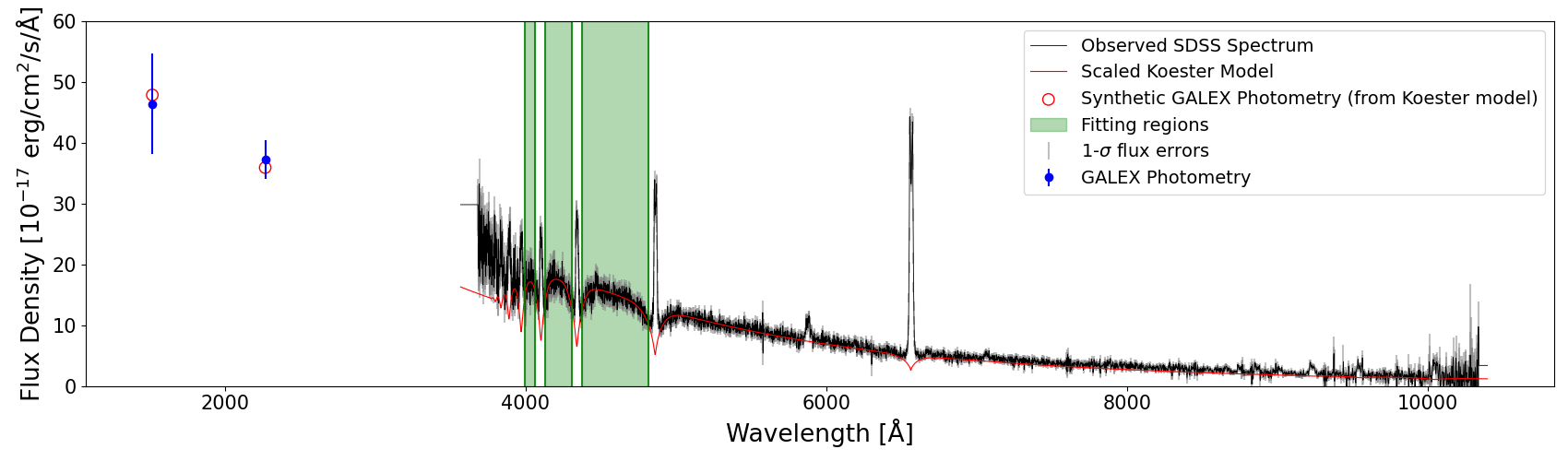}
    \caption{SDSS-V spectrum of GALEX\,J113749.9-200737 (MJD 60798) fitted with a \citet{koester2010} DA WD photospheric model, shown as an example of the adopted fitting procedure. The best-fitting model corresponds to $\log(g)$=8.25 and T$_{\rm eff}$=12500\,K.}
    \label{Fig.WD_fitting}
\end{figure*}

We visually inspected the SDSS-V spectra of all 32 eROSITA-selected high-likelihood PB candidates confirmed as CVs to search for signatures of WD photospheric absorption. The spectroscopic observing dates were compared with the long-term light curves of each system (see Sect.~\ref{subsec:long-term_LCs}) to ensure that the spectra were obtained during quiescence. In addition, we examined the spectra for indications of observations taken during or near an outburst. In outburst, CVs typically display a markedly bluer continuum relative to quiescence \citep[e.g.][]{Bailey_1980, Cathy_1990, Nogami_2004}, and as the accretion disc becomes optically thick, the Balmer emission cores can transition into broad absorption features that dominate the line profiles \citep[e.g.][]{Hessman_1984}.

In all 32 systems, the Balmer lines display broad absorption wings consistent with a WD photospheric origin, and the spectra are compatible with quiescent states. The SDSS-V spectra can thus be employed to estimate the WD parameters. We modelled the WD contribution to the observed spectra by fitting hydrogen-rich (DA) atmosphere models, deriving the effective temperature, $T_{\rm eff}$, surface gravity, $\log(g)$, and dilution factor $(R_{\rm WD}/d)^2$, where $R_{\rm WD}$ is the WD radius and $d$ is the distance of the system. Our fitting strategy follows the methodology described in \citet{Hernandez-Diaz_2026} and \citet{Daniela_2026}. Here, we provide a short summary.

We adopted a grid of pure-hydrogen, local thermodynamic equilibrium (LTE) atmosphere models by \citet{koester2010}. The grid spans $T_{\rm eff}=8000$–$30000$\,K, sampled in steps of 250\,K up to 20000\,K and 1000\,K at higher temperatures, and $\log(g)$=6.5 - 9.5 in increments of 0.25\,dex. The WD models were convolved with the Gaussian instrumental profiles of the SDSS-V spectra to reproduce their spectral resolution and resampled via cubic-spline interpolation to the wavelength grid of the observed spectra.

To break degeneracies in the optical fitting, we included near- and far-ultraviolet (NUV and FUV) photometry from GALEX in the fitting procedure. Synthetic photometry in the two GALEX bands was derived using the model spectra and the GALEX NUV and FUV filter transmission curves \citep[see e.g.][]{Casagrande}.

The spectral fitting was performed by minimising a global $\chi^2$ statistic that jointly incorporates the SDSS spectroscopic data and the GALEX photometry (see Eq.~1 in \citealt{Hernandez-Diaz_2026}). We restricted the fitting to wavelength regions where the WD dominates, excluding the red optical range and masking the Balmer emission cores, while including the absorption wings from the WD (see Fig.~\ref{Fig.WD_fitting}). Among the best-fitting models yielding comparably low total $\chi^2$ values, the adopted solution corresponds to the one best reproducing the GALEX UV photometry. For systems without available GALEX photometry, the fitting was conducted using only the optical spectrum.

Under the assumption that the continuum emission from the accretion disc and hot spot in PBs can be approximated by a power law, their combined contribution in the UV and blue optical spectrum is expected to be of the order of $\sim$10–20\% for PBs, and it can be negligible in some systems (see \citealt{Pala_2022} and \citealt{Neustroev_2023}). In addition, in CVs with low mass accretion rates, the boundary layer is expected to be optically thin and to radiate predominantly in hard X-rays with minor contribution in the UV (see e.g. \citealt{Narayan_1993}). 

Given the dominance of the WD at these wavelengths, the uncertain shape of the continuum emission from the low-accretion-rate disc (see e.g. \citealt{Neustroev_2026}), and the limited optical wavelength coverage of the SDSS-V spectra, we decided not to include an additional disc component into our modelling, thereby avoiding the introduction of additional complexity that is not well constrained by the available data. We also note that an accurate characterisation of WDs in CVs is generally best achieved using UV spectroscopy, as the SED of WDs in CVs peaks in the UV (e.g. \citealt{Pala_2017,Toloza_2023}). In absence of such measurements the parameters derived from our optical spectral fits can be used as approximate physical parameters of the WDs. 

To estimate statistical uncertainties on the derived $T_{\rm eff}$, $\log(g)$, and $(R_{\rm WD}/d)^2$, we adopted a Monte–Carlo approach. For each observed spectrum, we generated 100 synthetic realisations by perturbing the SDSS flux density values with Gaussian noise consistent with their $1\sigma$ uncertainties. When available, the GALEX FUV and NUV magnitudes were perturbed analogously using their respective $1\sigma$ uncertainties. Each resampled dataset was then fitted using a WD model grid restricted to $T_{\rm eff}$ within $\pm1000$\,K and $\log(g)$ within $\pm1$\,dex of their best-fitting estimates. To account for possible asymmetries in the resulting parameter distributions, the lower and upper uncertainties on $T_{\rm eff}$, $\log(g)$, and $(R_{\rm WD}/d)^2$ were computed as the differences between the best-fitting values and the 16th and 84th percentiles, respectively. In the case of symmetric (Gaussian) distributions, these correspond to a $1\sigma$ confidence interval. In a few cases, the resampling procedure did not lead to variations in the best-fitting model parameters, such that one or both of the 16th and 84th percentiles coincide with the best-fitting value. This situation can also arise when the best-fitting parameter lies at the boundary of the model grid. In these cases, we define the uncertainties in $T_{\rm eff}$ and $\log(g)$ as half the step size of the model grid.

We note that the system \textit{Gaia} DR3 6396267195026243840 was excluded from this analysis owing to the low S/N of its SDSS-V spectra. Additionally, for the magnetic systems in our sample that exhibit Zeeman splitting  (see Sect.~\ref{subsec:Zeeman_Splitting}), the derived parameters should be treated with caution. Although no evident cyclotron harmonic features are apparent in the continuum, a contribution from cyclotron emission cannot be excluded. In addition, Zeeman splitting distorts the Balmer absorption profiles. Consequently, the absorption wings from the WD could not be used in the spectral fitting. For this reason we fixed $\log(g)=8.5$ for these systems. In the case of V379\,Vir, literature estimates place $T_{\rm eff}\sim$10000–11000\,K (\citealt{Schmidt_2005}, \citealt{Suslikov_2025b}), whereas our modelling yields $\sim$14000\,K. Given this discrepancy, the magnetic nature of the system, and the absence of GALEX photometry to better constrain the fit, we excluded this object from this analysis.

The dilution factors, $(R_{WD}/d)^2$, obtained from our spectral fits were then used together with the distances from \citet{Bailer-Jones} to estimate $R_{\text{WD}}$. The mass of the WD, $M_{\text{WD}}$, was then obtained from the mass–radius relationship\footnote{\url{https://www.astro.umontreal.ca/~bergeron/CoolingModels/}} of \citet{Holberg_2006}, \citet{Tremblay_2011}. Finally, we derived the secular mass accretion rates onto the WD, $\langle\dot{M}\rangle$, from Eq.~1 of \citet{Townsley_2009}. We note that, if the accretion disc contributes to the blue optical spectrum, neglecting its contribution could result in an overestimation of $R_{\mathrm{WD}}$ and an underestimation of $M_{\mathrm{WD}}$ for some systems, such that the true $\langle\dot{M}\rangle$ may be lower than inferred.

The statistical uncertainties on $R_{\rm WD}$, $M_{\rm WD}$, and $\langle\dot{M}\rangle$ were estimated through Monte–Carlo error propagation. In this procedure, $T_{\rm eff}$, $(R_{WD}/d)^2$, and the \citet{Bailer-Jones} distances were randomly resampled 1000 times according to their uncertainties, and the parameters $R_{\rm WD}$, $M_{\rm WD}$, and $\dot{M}$ were recalculated at each realisation. The lower and upper uncertainties on $R_{\rm WD}$, $M_{\rm WD}$, and $\langle\dot{M}\rangle$ were derived from the 16th and 84th percentiles of their Monte-Carlo distributions relative to the best-fitting values. We note that for one spectrum of QZ\,Lib the resulting distribution for $R_{\rm WD}$ is strongly asymmetric, with the probability mass concentrated near the best-fit value, one tail significantly suppressed, and the opposite tail extending over a broader range. In this case, one of the percentile bounds lies beyond the best-fit value, preventing the definition of the corresponding uncertainty.

The measured $T_{\rm eff}$, $\log(g)$, $R_{\rm WD}$, $M_{\rm WD}$, and $\langle\dot{M}\rangle$ are available electronically at CDS (see Table~\ref{table:WDs_parameters} in Appendix~\ref{sec:WDs_parameters}). A flag indicates whether GALEX photometry was included in the spectral fitting. We discuss our results in Sect.~\ref{subsec:Teff_Mdot_Distributions}.

\subsection{Spectral energy distributions}
\label{subsec:SEDs}

We attempted to construct SEDs for the CVs in our sample using the multi-wavelength photometry described in Sect.~\ref{subsec:multiwavelength_phot}. 
Our CV sample consists of 32 systems exhibiting Balmer emission lines in their SDSS-V spectra (see Sect.~\ref{subsec:emission_lines}) and the 14 systems not observed in SDSS-V but displaying dwarf-nova outbursts in their long-term light curves (see Sect.~\ref{subsec:Outbursts}), where the AM\,CVn systems have been excluded (see Sect.~\ref{subsec:AM_CVns}). We additionally included EF\,Eridani and \textit{Gaia} DR3 2884625449040056704 in this analysis, as we identified transitions between high and low states typical of mCVs and measured $P_{\rm orb}$ (see Sect.~\ref{subsec:confirmation_PBs}, Table~\ref{sec:orbital_periods_results}, and Appendix~\ref{sec:special_systems_Plots}). Four systems lack sufficient photometric coverage for a SED analysis, and additionally seven systems lack WISE infrared photometry, required to constrain the contribution of the donor. The final sample for the SED analysis therefore consisted of 37 systems.

We performed the SED fitting using the Virtual Observatory SED Analyzer (VOSA, \citealt{Bayo_2008}), adopting a two-component fitting approach, that combines a WD atmosphere model (\citealt{koester2010}) and a low-mass star model for the donor (BT-Settl; \citealt{Caffau_2011}, \citealt{Allard_2012}). To constrain the WD component, we used the $T_{\mathrm{eff}}$ estimates obtained from our spectroscopic analysis when available (see Sect.~\ref{subsec:WD_model_fitting}), constraining the fit to within $\pm1000$ K around these values. The WD $\log(g)$ was allowed to vary over $7.25 \leq \log(g) \leq 8.75$ by default, unless our spectroscopic analysis favoured values outside this interval (see Sect.~\ref{subsec:WD_model_fitting}), in which case the allowed range was extended accordingly. We note, however, that a broadband SED does not provide reliable constraints on $\log(g)$, and we therefore do not report the inferred WD and donor $\log(g)$ values. Examples of SED fittings are presented in Appendix~\ref{sec:special_systems_Plots}. 

An approximate donor $T_{\rm eff}$ was inferred from the IR excess in the SED. The corresponding SpT of the donor was then inferred using the semi-empirical evolutionary tracks of \citet{Knigge_2011}. We note that this photometric inference should be treated with caution as the IR excess can also include contributions from the accretion disc in non-magnetic CVs or cyclotron emission in mCVs, potentially leading to an overestimation of the donor contribution and, consequently, an earlier inferred SpT. Thus, the true SpT of the donor may be later than indicated by our SED fitting. Clearly, a final confirmation of the PB status requires a spectroscopic detection of the typical (molecular) features of a very late-type star or brown dwarf. The SpT inferred from the SED should only be regarded as conclusive for PB status when combined with additional evidence, such as X-ray emission and multi-wavelength properties consistent with bona fide PBs, and a $P_{\rm orb}$ determination close to the period minimum (see \citealt{Daniela_2024}).

To assess the impact of a possible underestimation of the WD $T_{\rm eff}$ estimates derived from our spectroscopic analysis (see Sect.~\ref{subsec:WD_model_fitting}), we repeated the SED fitting for a subset of systems after fixing the WD $T_{\rm eff}$ to values 1000 and 2000\,K higher than the best-fit value. The inferred donor $T_{\rm eff}$ values changed by a mean absolute value of $\sim$200\,K, with no systematic trend. Since we infer that most of our PB candidates host T dwarf donors (see Sect.~\ref{subsec:confirmation_PBs} and Table~\ref{sec:SED_Results}), these potential variations do not affect their classification as strong PB candidates. 

The $T_{\rm eff}$ and $\log g$ values of the best-fitting models, as well as the inferred SpTs of the donors are available electronically at CDS (see Table~\ref{table:SED_Results} in Appendix~\ref{sec:SED_Results}). We discuss our results in Sect.~\ref{subsec:confirmation_PBs}.

\subsection{Balmer decrement analysis}
\label{subsec:Balmer_decrements}

In \cite{Hernandez-Diaz_2026}, we introduced the Balmer decrements as a new diagnostic to identify PBs. We measured the emission-line luminosities of H$\alpha$, H$\beta$, and H$\gamma$ (hereafter $L_{\mathrm{H}{\alpha}}$, $L_{\mathrm{H}{\beta}}$, and $L_{\mathrm{H}{\gamma}}$), and demonstrated that the $L_{\mathrm{H}{\gamma}}/L_{\mathrm{H}{\beta}}$ versus $L_{\mathrm{H}{\alpha}}/L_{\mathrm{H}{\beta}}$ diagram represents an effective means for separating PBs from short-period pre-bouncers. Specifically, PBs were found to exhibit systematically steeper Balmer decrements, that is, $\text{H}{\alpha}/\text{H}{\beta}>1$ and $\text{H}{\gamma}/\text{H}{\beta}<1$, compared to pre-bouncers.

To quantitatively separate both populations, we trained a linear logistic regression model (\citealt{Berkson_1944}, \citealt{Cox_1958}) in the Balmer decrement parameter space. This model assigns to each system a probability of being a PB, $P_{\text{PB}}$, based on the observed Balmer ratios. Using the training sample of \citet{Hernandez-Diaz_2026}, we found a classification threshold of $P_{\rm PB}=0.535$, above which systems are classified as PBs.

In this work, we adopted the methodology described by  \citet{Hernandez-Diaz_2026}. We subtracted the best-fitting DA WD atmosphere model (see Sect.~\ref{subsec:WD_model_fitting}) from the observed spectrum to correct for the photospheric Balmer absorption of the WD. We then measured the emission-line fluxes of H$\alpha$, H$\beta$, and H$\gamma$, and converted these fluxes into line luminosities using the distances from \citet{Bailer-Jones}.

In this analysis, we excluded the four mCVs identified in our sample (see Sect.~\ref{subsec:Zeeman_Splitting}). In magnetic systems, the magnetically controlled accretion flow is expected to suppress the formation of a disc and,  therefore, the systems do not satisfy the assumption of disc-dominated emission underlying our diagnostic (see \citealt{Hernandez-Diaz_2026}). For another two systems (GALEX\,J035124.8-092742 and \textit{Gaia} DR3 6396267195026243840) the Balmer lines could not be reliably measured in some or all the available spectra due to the absence of emission in H$\beta$, H$\gamma$, or both, or because these regions of the spectrum are dominated by noise.

We note that spectra obtained during or close to an outburst can lead to biased Balmer decrements compared to quiescence (see \citealt{Hernandez-Diaz_2026} and references therein). However, as mentioned in Sect.~\ref{subsec:WD_model_fitting}, all the SDSS-V spectra analysed in this work were obtained during quiescence.

The measured line fluxes, line luminosities, Balmer decrements, and associated $P_{\text{PB}}$ values are available electronically at CDS (see Table~\ref{sec:balmer_decrements_results} in Appendix~\ref{table:Balmer_Decrements}). A discussion of our results is presented in Sect.~\ref{subsec:BD_discussion}.

\subsection{Zeeman splitting analysis in magnetic CVs}
\label{subsec:Zeeman_Splitting}

Four eROSITA-selected high-likelihood PB candidates exhibit Balmer Zeeman splitting in their SDSS-V spectra (see Fig.~\ref{fig:detached_PBs_panel} in Appendix~\ref{sec:spectra_semi-detached}). Specifically, CP\,Tuc and GALEX\,J033833.9-332802 show Zeeman splitting in both H$\alpha$ and H$\beta$. However, in H$\beta$, the central Zeeman component is strongly affected by superimposed emission, preventing a reliable modelling of the underlying absorption profile. Thus, we used only H$\alpha$ to estimate the mean photospheric magnetic field strength of the WD, $\langle B \rangle$, in these systems. Finally, 2QZ\,J130249.7+014919 and V379\,Vir display noticeable Zeeman splitting only in H$\beta$, which we used to determine $\langle B \rangle$.

\begin{figure}
	\centering
	\includegraphics[width=0.95\linewidth]{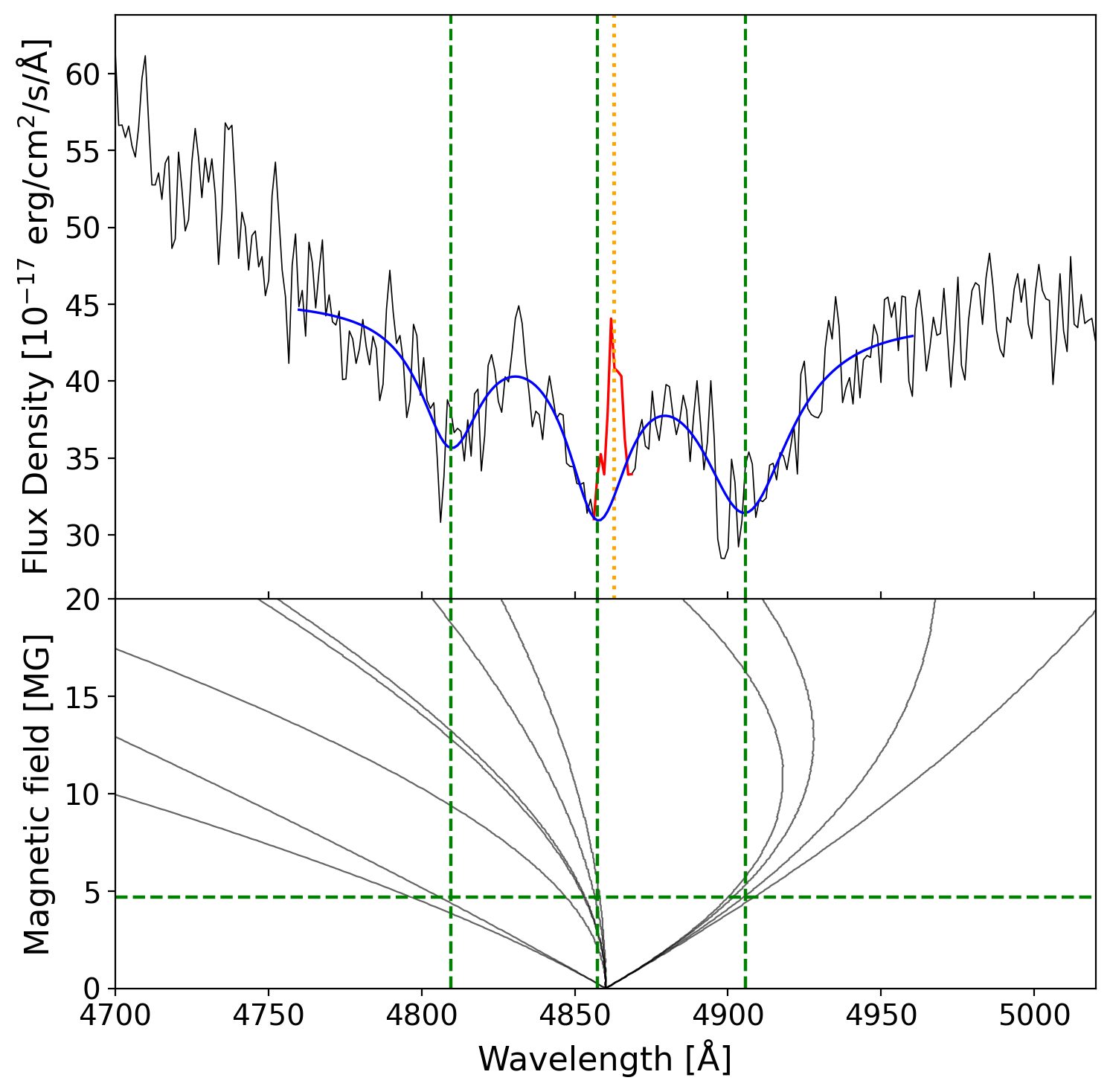}
	\caption{SDSS-V spectrum of V379\,Vir (observed at MJD 60411) around the H$\beta$ line, as an example of a system showing Zeeman splitting. Top panel: Observed spectrum (black) with the best-fitting set of three Lorentzian components plus a low-order polynomial continuum overplotted in blue (see Eq.~\ref{eq:model_Zeeman_Splitting}). The orange dotted line marks the vacuum rest wavelength of H$\beta$ at 4862.721\AA. The region of the emission component shown in red is excluded from the fitting. The green dashed vertical lines indicate the centroid positions obtained from the fit, $\mu_{\rm i}^{\rm fit}$. Bottom panel: Theoretical Zeeman transition branches as a function of the magnetic field strength, $\lambda(B)$ (black curves; see text). The horizontal green dashed line marks the best-fitting magnetic field value obtained from the $\chi^2$-minimisation (see Eq.~\ref{eq:chi2_Zeeman_Splitting}). The corresponding analyses for CP\,Tuc, GALEX\,J033833.9-332802, and 2QZ\,J130249.7+014919 are shown in Fig.~\ref{fig:Zeeman_Splitting_Plots} in Appendix~\ref{sec:Zeeman_splitting_results}.}
	\label{Fig.Zeeman_Splitting}
\end{figure}

Similarly to the approach in \citet{Suslikov_2025}, the Zeeman splitting of the Balmer lines was modelled by fitting the observed line profile with a set of three Lorentzian components plus a low-order polynomial continuum (see Fig.~\ref{Fig.Zeeman_Splitting} for an example). This is expressed as

\begin{equation}
F(\lambda) = P_n(\lambda) + \sum_{i=1}^{3}
\frac{A^{\mathrm{fit}}_{i}}{1 + \left( \frac{\lambda - \mu^{\mathrm{fit}}_{i}}{\gamma^{\mathrm{fit}}_{i}} \right)^2}
\label{eq:model_Zeeman_Splitting}
\end{equation}

\noindent where $P_n(\lambda)$ is an $n$-th-order polynomial used to describe the continuum, and $A^{\mathrm{fit}}_{i}$, $\mu^{\mathrm{fit}}_{i}$, and $\gamma^{\mathrm{fit}}_{i}$ are the amplitude, centroid wavelength, and width of each Lorentzian component, respectively.

The fit was performed using nonlinear least squares minimisation using the Trust Region Reflective algorithm (\citealt{Branch_1999}, \citealt{Nocedal_2006}) as implemented in the \textsl{Scipy} Python package. In systems where an emission component is present in the line core, the affected wavelength region was masked prior to the fit to prevent the emission feature from biasing the centroid determination of the absorption Zeeman components (see Fig.~\ref{Fig.Zeeman_Splitting}).

We computed the wavelengths of the allowed Zeeman transitions for H$\alpha$ and H$\beta$ on a grid of magnetic field strengths ranging from 0 to 50\,MG in steps of 0.05\,MG. The calculations were performed using the code described by \citet{Schimeczek_2014}, which provides accurate transition wavelengths and oscillator strengths beyond the linear Zeeman regime, up to $B\sim10^{7}$\,MG.

For each magnetic field value $B_k$ in the grid, the set of allowed transitions $\{\lambda_{j}(B_{k}), f_{j}(B_{k})\}$ was extracted, where $\lambda_{j}(B_{k})$ and $f_{j}(B_{k})$ denote the wavelength and oscillator strength of  transition $j$, respectively. For each of the three observed Zeeman components, transitions within a wavelength interval of width $\gamma_{i}^{\mathrm{fit}}$ centred on the fitted centroid $\mu_{i}^{\mathrm{fit}}$ were selected, and the corresponding theoretically expected centroid at that $B_{k}$, $\mu_{i}^{\mathrm{theo}}(B_{k})$, was computed as their oscillator-strength weighted mean,

\begin{equation}
\mu_{i}^{\mathrm{theo}}(B_{k})
=
\frac{\sum_{j} f_{j}(B_{k})\,\lambda_{j}(B_{k})}
{\sum_{j} f_{j}(B_{k})}
\label{eq:theo_centroid}
\end{equation}

\noindent The best-fitting magnetic field strength was then determined by minimising the $\chi^2$ statistic between the fitted Lorentzian centroids $\mu_{i}^{\mathrm{fit}}$ and the theoretical centroids $\mu_i^{\mathrm{theo}}(B_k)$,

\begin{equation}
\chi^{2}(B_{k}) =
\sum_{i=1}^{3}
\frac{\left( \mu_i^{\mathrm{fit}} - \mu_{i}^{\mathrm{theo}}(B_{k}) \right)^{2}}{\sigma_{\mu_i^{\mathrm{fit}}}^{2}},
\label{eq:chi2_Zeeman_Splitting}
\end{equation}

\noindent where $\sigma_{\mu_i^{\mathrm{fit}}}$ are the 1$\sigma$ uncertainties associated with $\mu_i^{\mathrm{fit}}$. The best-fitting magnetic field is then given by  

\begin{equation}
\langle B \rangle = \arg\min_{B_{k}} \chi^{2}(B_{k}).
\label{eq:best_B}
\end{equation}

\noindent The uncertainty in the derived $\langle B \rangle$ was estimated using a Monte Carlo approach. We extracted the $3 \times 3$ submatrix corresponding to the fitted centroid parameters $\mu_i^{\mathrm{fit}}$ from the full covariance matrix obtained in the Zeeman splitting profile fitting (see Eq.~\ref{eq:model_Zeeman_Splitting}). This marginalised covariance matrix was then used to draw new centroid values from a multivariate normal distribution at each Monte Carlo iteration. For each realisation, the centroids $\mu_{i}^{\mathrm{theo}}(B_{k})$ were recomputed across the magnetic field grid and the $\chi^2$-minimisation was repeated. This procedure was performed 1000 times, yielding a distribution of $\langle B \rangle$. The lower and upper uncertainties on $\langle B \rangle$ were defined as the differences between the best-fit value and the 16th and 84th percentiles.

In Table~\ref{table:magnetic fields} (see Appendix~\ref{sec:Zeeman_splitting_results}), we present the $\langle B \rangle$ values derived from the Zeeman splitting analysis (see also Fig.~\ref{fig:Zeeman_Splitting_Plots}). We note that since we use only one line to constrain $\langle B \rangle$, there is a higher risk of degeneracy, that is, different $\langle B \rangle$ values may reproduce the observed centroid separations with comparable $\chi^{2}(B_{k})$ values. To assess this possibility, we inspected the $\chi^{2}(B_{k})$ curves and searched for secondary local minima with $\chi^{2}$ values similar to that of the global minimum. We consider a secondary minimum to be statistically comparable to the global minimum if $\Delta \chi^2 = \chi^2(B_k) - \chi^2_{\min} \le 4$, which corresponds to the $2\sigma$ confidence level for one free parameter. In all but one case the solution is unique. For 2QZ\,J130249.7+014919, however, two distinct $\langle B \rangle$ values yield comparably low $\chi^{2}$ values. We therefore report both solutions for this system.

CP\,Tuc and V379\,Vir are known PBs (\citealt{Daniela_2026}, \citealt{Farihi_2008}) with previously reported magnetic field estimates. Our $\langle B \rangle$ measurement for CP\,Tuc is very close to the most probable photospheric value of $10\,$MG reported by \citet{Beuermann_2007}, while for V379\,Vir, our measurement lies within the range inferred from the rotational modulation of $\langle B \rangle$ (\citealt{Suslikov_2025}). 

\section{Discussion}
\label{Discussion}

\subsection{Period detections and double-humped modulations in TESS light curves}
\label{subsec:periods}

Among the systems with TESS two-minute cadence light curves in quiescence, we report two new $P_{\rm orb}$ detections. In addition, we detected $P_{\rm orb}$ for four systems with previously reported values in the literature, finding consistent results (see Table~\ref{table:orbital_periods}). We also identify two tentative period detections with their $\text{S/N}_{\text{PSD}}$ values lying below the reliability threshold of $\left(\text{S/N}_{\text{PSD}}\right)_{\text{min}} = 0.004$ (see Sect.~\ref{subsec:Porb}). Notably, the periodic signal detected for GALEX\,J022318.3-431701 is independently recovered using three methods, supporting its credibility despite the low $\text{S/N}_{\text{PSD}}$.

This low rate of period detections in the TESS light curves was expected. As shown by \cite{Hernandez-Diaz_2025}, period detections become increasingly unreliable for systems with TESS magnitudes $\gtrsim 17$\,mag, where noise begins to dominate. Most systems in our sample are fainter than this limit, with typical TESS magnitudes $\gtrsim 18$\,mag.

All detected signals are interpreted as $P_{\rm orb}$, except for GALEX\,J022318.3-431701, where the $39.539 \pm 0.014$ min periodicity likely corresponds to $\frac{1}{2}P_{\rm orb}$. SDSS-V spectroscopy rules out an AM\,CVn classification for this system, as it shows clear Balmer emission, suggesting that this short period corresponds to the second harmonic of $P_{\rm orb}$, while the fundamental orbital signal is likely suppressed by noise. Alternatively, the detected signal could correspond to the WD spin period. Double-humped variability, in which the brightness modulation at $\frac{1}{2}P_{\rm orb}$ dominates over the orbital phase, is common in CVs and can arise from several mechanisms. In non-magnetic CVs, double-humped brightness modulations can originate from ellipsoidal variability of the secondary (\citealt{Bochkarev_1979}, \citealt{Wilson_2006}), partial eclipses between the accretion disc and the secondary at high inclination (\citealt{Cynthia_1999}), or orbital changes in the projected area of an elongated, continuously visible hotspot in an optically thin accretion disc (\citealt{Skidmore_2000}). Of particular importance for short-period CVs is the formation of two spiral density waves in the outer parts of the accretion disc when its radius reaches the 2:1 resonance with the donor \citep{Kunze_2005}, generally expected to occur in systems with low mass ratios ($q \lesssim 0.1$). This mechanism can also produce double-humped brightness modulations in the observed light curve \citep{Aviles_2010, zharikov2013, Amantayeva_2021}. In mCVs, double-humped brightness modulations can arise from two-pole accretion onto the WD (\citealt{Ferrario_1989}), from cyclotron beaming produced in the accretion regions of the WD (\citealt{Schwarm_2017}), or from the aforementioned ellipsoidal variability of the donor. In our sample, double-humped brightness modulations are observed in GALEX\,J125751.4-283015, BW\,Scl, USNO-A2.0 0525-17852977, and \textit{Gaia} DR3 2884625449040056704.

\subsection{WD effective temperatures and secular mass-accretion rates in PB Candidates}
\label{subsec:Teff_Mdot_Distributions}

\begin{figure*}
	\centering
	\includegraphics[width=0.45\linewidth]{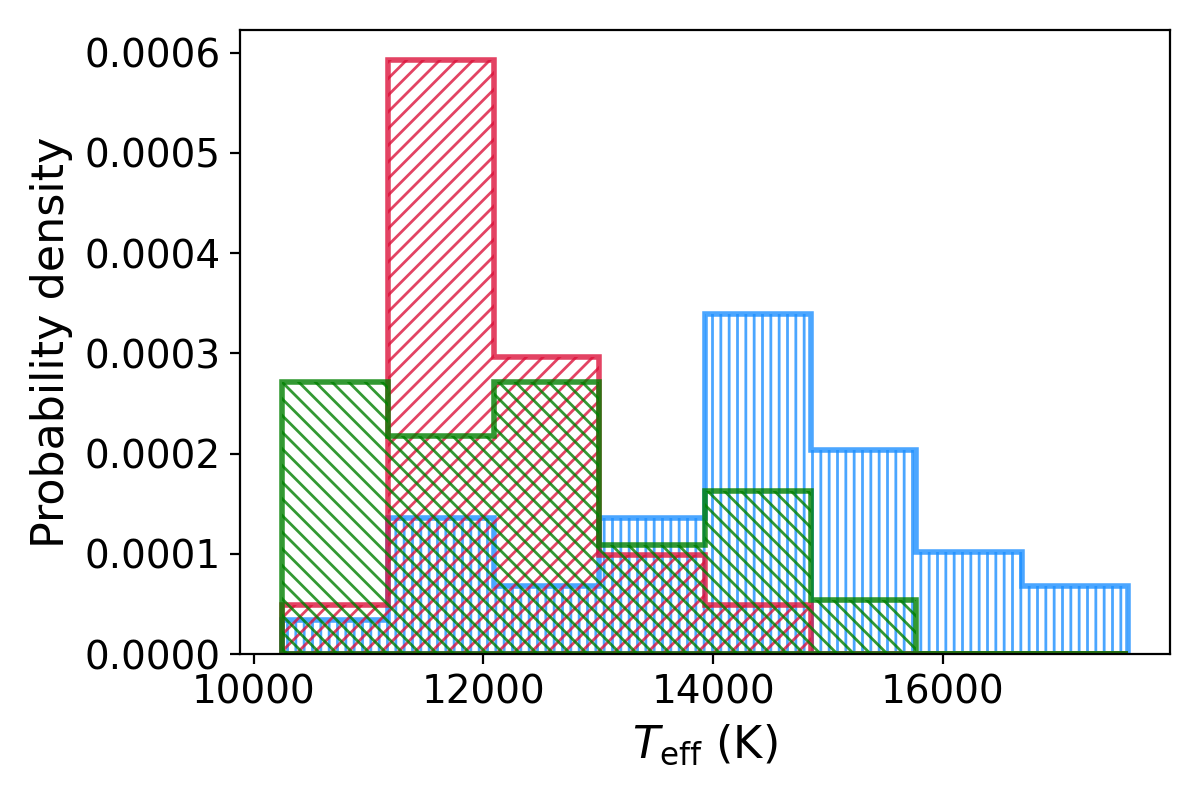}
    \includegraphics[width=0.45\linewidth]{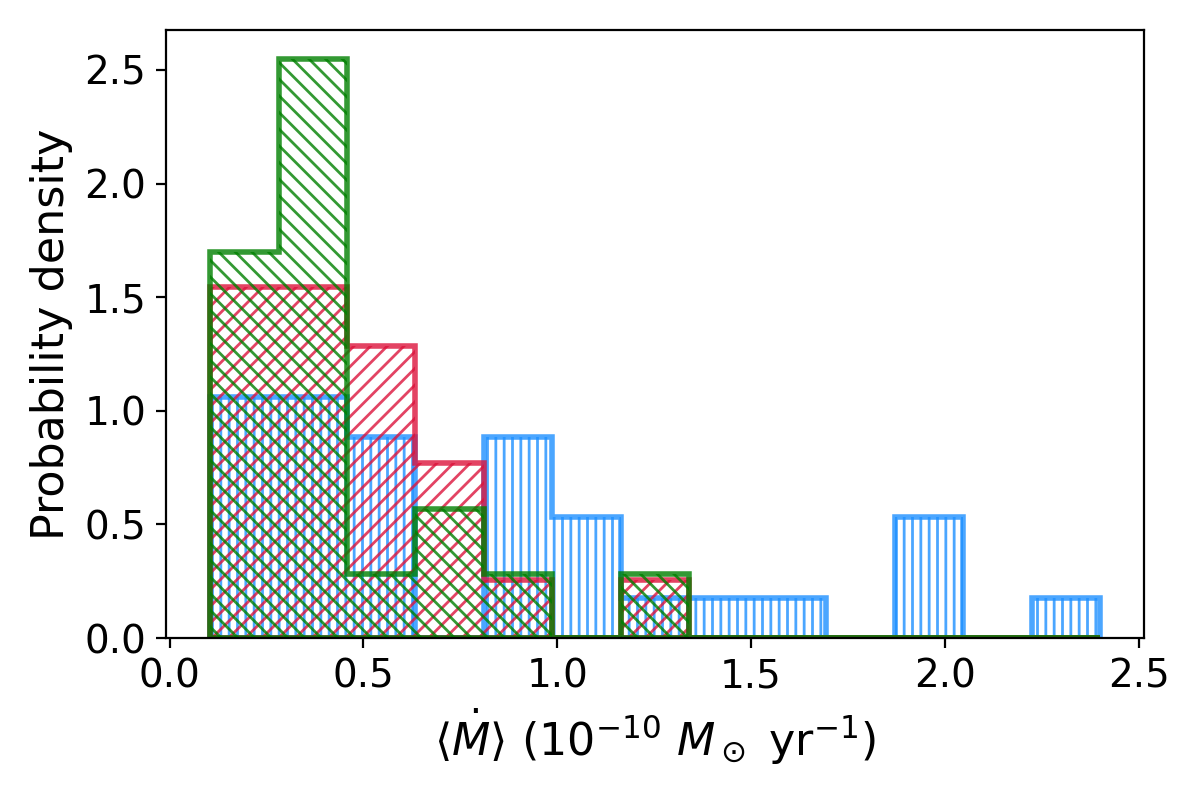}
	\caption{Distributions of measured WD $T_{\rm eff}$ (left) and $\langle\dot{M}\rangle$ (right). Red and green histograms correspond to the confirmed PBs and PB candidates in our sample, respectively. Blue histograms show the short-period ($P_{\rm orb}<2\,h$) CVs from \citet{Pala_2022}. All distributions are normalised to unit area, ensuring that each distribution represents an empirical probability density function and facilitating the comparison between samples.}
	\label{Fig.WD_parameters_Distributions}
\end{figure*}

After the common-envelope phase, CV progenitors remain detached for $\sim0.1-10$ Gyr. By the time angular momentum losses bring the donor into Roche-lobe overflow, the WD has cooled to $T_{\rm eff}\approx4500-5000\,K$ (\citealt{Althaus_1997}, \citealt{Salaris_2000}). The onset of mass-transfer then reheats the WD (Sion 1991). Measurements of WD $T_{\rm eff}$ in CVs during quiescence, when the UV emission is dominated by the WD, generally find $T_{\rm eff}>10000\,K$ (see e.g. \citealt{Sion_1999}, \citealt{Townsley_2009}, \citealt{Pala_2022}), confirming that WDs in CVs are hotter than isolated WDs of similar age.

This reheating of the WD is primarily driven by compressional heating. As newly accreted material compresses the underlying layers, the release of gravitational energy within the WD's interior is converted into thermal energy. This internal energy gradually diffuses outward on the thermal time-scale of the WD envelope ($10^{3}-10^{5}$ yr), setting the observed $T_{\rm eff}$ (\citealt{Sion_1995}, \citealt{Townsley_2009}). Consequently, $T_{\rm eff}$ provides a diagnostic of the secular mass-accretion rate $\langle\dot{M}\rangle$, time-averaged over the thermal time-scale of the WD (\citealt{Townsley_2003}, \citealt{Townsley_2004}, \citealt{Townsley_2009}).

As described in Sect.~\ref{subsec:WD_model_fitting}, in this work we modelled the WD contribution using DA atmosphere models fitted to SDSS-V spectra and GALEX UV photometry. Subsequently, assuming a WD mass-radius relation (\citealt{Holberg_2006}, \citealt{Tremblay_2011}), we constrained $T_{\rm eff}$ and $\langle\dot{M}\rangle$. In Fig.~\ref{Fig.WD_parameters_Distributions}, we present the distributions of the measured $T_{\rm eff}$ and $\langle\dot{M}\rangle$ values for our sample of 30 eROSITA-selected high-likelihood PB candidates with SDSS-V spectra characteristic of a CV, as indicated by the Balmer emission, for which we modelled the WD contribution.  
\citet{Pala_2022} determined $T_{\rm eff}$ and $\langle\dot{M}\rangle$ for a sample of 43 CVs using UV spectroscopy from the Hubble Space Telescope (HST), including many short-period CVs (with 7 of them being known PBs). To place our measurements in context, we include their results for the 34 systems with $P_{\rm orb}<2$\,h in Fig.~\ref{Fig.WD_parameters_Distributions}.  
We excluded CRTS\,J153817.3+512338 from the sample of \citealt{Pala_2022}, as it is an outlier with a much hotter WD compared to other CVs at similar $P_{\rm orb}$ (see \citealt{Pala_2017}).

As shown in Fig.~\ref{Fig.WD_parameters_Distributions}, the systems in our sample exhibit $T_{\rm eff}$ and $\langle\dot{M}\rangle$ values that are generally lower than those from \citet{Pala_2022}. Four PBs from the sample of \citet{Pala_2022} are also included in our sample. For three of these systems (HV\,Vir, LP\,731-60, and QZ\,Lib), our $T_{\rm eff}$ estimates agree closely with the values of \citet{Pala_2022}, with an average absolute difference of only $\sim$130\,K. For BW\,Scl, our estimate of 14000\,K is 1145\,K lower than that of \citet{Pala_2022}, but is in good agreement with the value of 14250\,K reported by \citet{Neustroev_2023}.

Notably, the distribution of our PB candidates show $T_{\rm eff}$ and $\langle\dot{M}\rangle$ values peaking at similar values to those of already confirmed PBs. This behaviour is consistent with the idea that a significant fraction of our PB candidates are indeed PBs and not short-period pre-bouncers. 

\subsection{PB probabilities from Balmer decrements}
\label{subsec:BD_discussion}

As discussed by \citet{Hernandez-Diaz_2026}, the steeper Balmer decrements observed in PBs reflect systematically different physical conditions within their accretion discs compared to pre-bounce CVs. PBs are characterised by very low mass accretion rates, leading to cool and predominantly optically thin discs (see also \citealt{Neustroev_2026}). \citet{Hernandez-Diaz_2026} also found evidence of a stronger radial stratification in the physical conditions of the discs of PBs. The Balmer decrements, therefore, probe the different line-forming regions of the disc and provide an indirect tracer of the evolutionary state of a short-period CV.

\begin{figure}
	\centering
	\includegraphics[width=0.95\linewidth]{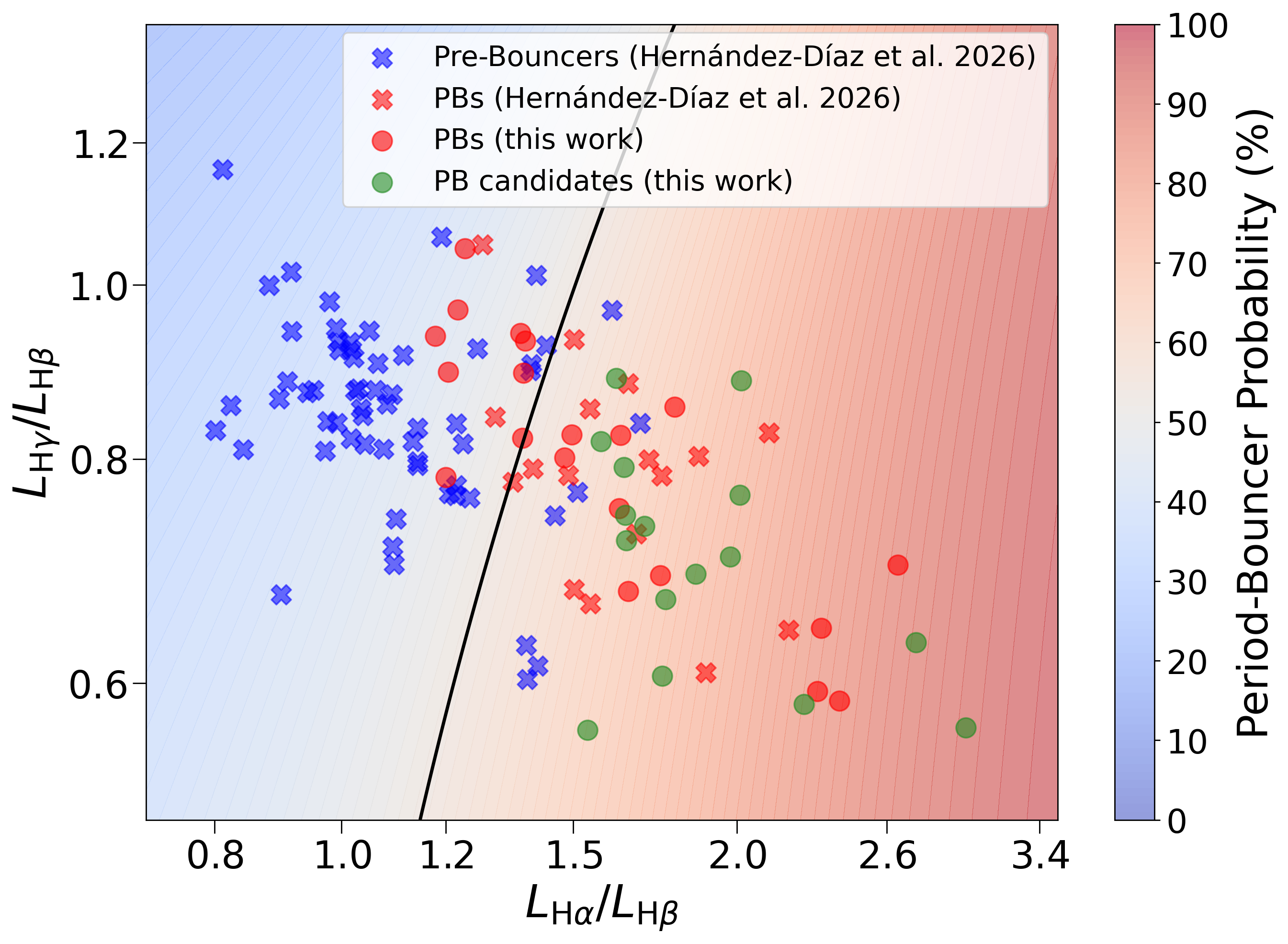}
	\caption{Diagnostic diagram of observed $L_{\mathrm{H}{\gamma}}/L_{\mathrm{H}{\beta}}$ ratio versus $L_{\mathrm{H}{\alpha}}/L_{\mathrm{H}{\beta}}$ ratio for previously confirmed and candidate PBs from our sample (red and green circles, respectively). The pre-bounce and period-bounce CVs from \citet{Hernandez-Diaz_2026} --which define the decision boundary at $P_{\text{PB}} = 0.535$ according to our linear logistic regression model-- are included for reference. The background colour scale represents the predicted probability that a system is a PB.}
	\label{Fig.Balmer-Decrements}
\end{figure}

In Fig.~\ref{Fig.Balmer-Decrements}, we present the measured Balmer decrements for our sample of 26 eROSITA-selected high-likelihood PB candidates (see Sect.~\ref{subsec:Balmer_decrements}), together with the reference samples of pre-bounce and period-bounce CVs from \citet{Hernandez-Diaz_2026}. Notably, all 14 eROSITA-selected high-likelihood PB candidates analysed here (green dots in Fig.~\ref{Fig.Balmer-Decrements}) are classified as PBs by our linear logistic regression model. They exhibit pronouncedly steep Balmer decrements, and their $P_{\text{PB}}$ values lie well above the adopted classification threshold at $P_{\text{PB}} = 0.535$, placing them firmly within the PB region of the diagnostic diagram. 

On the other hand, many of the 12 previously confirmed PBs that are part of our SDSS-V sample display comparatively flatter decrements, with some of them being located in the non-PB side of the Balmer decrement space. The lowest probability by our linear logistic regression model for a previously confirmed PB from the SDSS sample of this work that is misclassified as a pre-bounce system is $P_{\text{PB}}=43\%$ for V406\,Vir. Most of the remaining previously confirmed PBs misclassified by the model have $P_{\text{PB}}\gtrsim48\%$ and lie close to the decision threshold. 
This provides evidence for the presence of a transition region characterised by partial overlap between pre-bounce and period-bounce CVs, and suggests that the current population of confirmed PBs may be biased toward the less evolved post-bounce systems. As the population of confirmed PBs grows, the diagnostic diagram shown in Fig.~\ref{Fig.Balmer-Decrements} can be refined and reassessed in greater detail.

\subsection{Evolutionary status of PB candidates}
\label{subsec:confirmation_PBs}

As explained in Sect.~\ref{Introduction} and at the beginning of Sect.~\ref{Analysis}, 
we adopt the three-fold strategy described by \citet{Daniela_2026} to confirm our eROSITA-selected high-likelihood PB candidates: (i) confirmation of the system as a CV (Sect.~\ref{subsec:Confirmation_as_CVs} and Table~\ref{table:Master}), (ii) determination of $P_{\rm orb}$ near the period minimum at $\sim$80 min (Sect.~\ref{subsec:Porb} and Table~\ref{table:orbital_periods}), and (iii) photometric constraint of a very late-type donor (Sect.~\ref{subsec:SEDs} and Table~\ref{table:SED_Results}). Many of the high-likelihood PB candidates presented here satisfy two of these criteria and are further supported by additional diagnostics, namely the WD $T_{\rm eff}$ or the Balmer decrements (see Sect.~\ref{subsec:Teff_Mdot_Distributions} and Sect.~\ref{subsec:BD_discussion}). Nevertheless, we have not formally confirmed any new PB in this work, with the lack of $P_{\text{orb}}$ determinations standing as the primary limiting factor. In the following, we present a detailed assessment of our PB candidates in the context of these confirmation criteria.

Two systems that have not been observed in SDSS-V, USNO-A2.0 0525-17852977 and \textit{Gaia} DR3 2884625449040056704, likely satisfy the three core criteria adopted for PB confirmation; however, we did not classify them as confirmed PBs in this work. USNO-A2.0 0525-17852977 lies in a crowded field, and its 2MASS and WISE photometry are likely contaminated (\citealt{Favieres_2024}), making the SED-based donor SpT unreliable. For \textit{Gaia} DR3 2884625449040056704, there is no confirmation of its CV nature through the identification of Balmer emission lines or outbursts thus far. Nevertheless, its long-term light curve shows a transition to a high state at $\sim$MJD 58700 followed by a return to a low state at $\sim$MJD 59800, indicative of a mCV. We additionally measured $P_{\text{orb}} = 82.3059\pm0.0051$\,min from its TESS light curves, while the SED reveals an IR excess, consistent with a very cool donor with $T_{\rm eff}\lesssim1500$\,K. Overall, its properties resemble those of EF Eridani, a short-period polar also exhibiting high and low states (see e.g. \citealt{Schwope_2010}). For both systems, we suggest follow-up observations to further constrain their donor properties. In Appendix~\ref{sec:special_systems_Plots}, we present the Lomb–Scargle periodograms of their TESS light curves used to determine $P_{\rm orb}$, the phase-folded TESS light curves, their combined long-term light curves, and the binary SED fits.

Our spectroscopic analysis of 17 new CVs observed with SDSS-V (see Sect.~\ref{subsec:Teff_Mdot_Distributions} and Sect.~\ref{subsec:BD_discussion}) indicates that they are all consistent with being PBs. Notably, we conducted the SED analysis (see Sect.~\ref{subsec:SEDs}) for 11 of these new CVs, yielding donor $T_{\rm eff}$ below $1600$\,K in 10 cases, corresponding to substellar donors of SpT T (see Table~\ref{table:SED_Results}). For the remaining system, \textit{Gaia} DR3 914295050082457600, we inferred a donor with $T_{\rm eff}\lesssim 2350$\,K, corresponding to a SpT M8 or later, that is, the donor might be substellar in this case as well. We also performed SED modelling for five additional PB candidates not observed in SDSS-V but exhibiting dwarf-nova outbursts in their long-term light curves. In all cases, we derived donor temperatures below $1800$\,K, consistent with substellar donors of spectral type L4 or later. No $P_{\rm orb}$ values could be measured for these systems, and they are therefore not formally confirmed as PBs. Future follow-up observations will enable their confirmation, potentially leading to an increase of the population of confirmed PBs by $\sim 50\%$.

\subsection{PBs possibly approaching detachment}
\label{subsec:PBs_detached}

From a visual inspection of the SDSS-V spectra, we identified five eROSITA-selected high-likelihood PB candidates exhibiting extremely weak Balmer emission lines, in some cases limited to H$\alpha$, three of which are new CVs (see Fig.~\ref{fig:detached_PBs_panel} in Appendix~\ref{sec:spectra_semi-detached}). This indicates that these systems have very low mass-transfer rates, raising the possibility that these objects are PBs approaching detachment. Notably, two of these systems, V379\,Vir and CP\,Tuc, have been previously classified as PBs (\citealt{Farihi_2008}, \citealt{Daniela_2026}).

As proposed by \citet{Schreiber_2023} and \citet{Schreiber_2025}, the observed scarcity of PBs could be explained by the late appearance of magnetic fields in the WDs of evolved CVs. After the  field diffuses to the surface of the WD, it can couple to that of the donor, leading to synchronisation torques that transfer spin angular momentum from the WD back to the orbit. This angular momentum transfer would increase the orbital separation and cause the system to detach, interrupting mass-transfer for several gigayears. As a result, many PBs that would otherwise be observed as accreting systems may instead spend most of their evolution beyond the minimum orbital period in a detached state.

The five systems with very weak Balmer emission lines identified in this work are therefore promising candidates for nearly detached PBs. Significantly, 
for four of them the WD is magnetic, as indicated by the Zeeman splitting observed in H$\alpha$ and H$\beta$ (see Sect.~\ref{subsec:Zeeman_Splitting}), and their measured mean magnetic field strengths ($\langle B \rangle \approx 5$--15\,MG; see Table~\ref{table:magnetic fields}) place them at the lower end of the $\langle B \rangle$ distribution of known polars (see \citealt{Schwope_2025}). For the remaining system, the presence of a weaker magnetic field cannot be ruled out. As discussed by \citet{Schreiber_2023}, even a relatively modest field strength of the order of $\sim$1 MG is sufficient to produce an extended detached phase. 

For V379\,Vir and CP\,Tuc, X-ray observations have probed direct accretion onto the WD (see \citealt{Stelzer_2017}, \citealt{Ramsay_1999}, and \citealt{Silva_2013}). In addition, Doppler tomography of V379\,Vir indicates that at least part of the H$\alpha$ emission arises within the accretion stream (see \citealt{Suslikov_2025}). For the three remaining systems identified in this work, follow-up X-ray observations and time-resolved spectroscopy are required to determine whether these systems are already detached. In particular, these observations would help to distinguish between ongoing accretion onto the WD via Roche-lobe overflow of the donor, accretion from the wind of a detached donor, and emission dominated by the irradiation of the donor.

We also examined the long-term light curves of these systems to search whether they exhibit the high and low states characteristic of mCVs (see \citealt{Livio_1994}, \citealt{King_1998}; \citealt{Kafka_2005}). None of the systems display such transitions. Therefore, we cannot rule out the possibility that they are mCVs currently residing in an extended low state. In such a scenario, the mass-transfer rate would be temporarily suppressed, rather than reflecting a secular evolution toward detachment as described above. 

Finally, although four systems clearly host magnetic WDs, the evolutionary origin of their magnetic fields remains uncertain. The fields could have formed before the systems reached the orbital period minimum, implying that they evolved as mCVs throughout their secular evolution.

\section{Summary and conclusions}
\label{Summary}

We present an analysis of optical spectroscopic and time-domain follow-up observations for a subsample of 213 eROSITA-selected high-likelihood PB candidates. The sample includes 19 previously known PBs, which serve as benchmarks for evaluating the remaining candidates. They will also provide a reference for future follow-up studies.

We confirm 24 new CVs through the identification of Balmer emission lines in SDSS-V spectra (18 cases) or outburst activity in long-term light curves (8 cases) (Sect.~\ref{subsec:Confirmation_as_CVs}). Notably, only 2 out of the 18 newly confirmed CVs with SDSS-V optical spectra exhibit outbursts in their long-term light curves. This suggests that many PBs undergoing long quiescent phases have potentially eluded detection in variability-based CV searches that rely on outburst identification. This highlights the efficiency of our X-ray–based selection method in uncovering new CVs. 

Among the 78 eROSITA-selected high-likelihood PB candidates observed in SDSS-V, 32 systems exhibit Balmer emission (see Sect.~\ref{subsec:emission_lines}) and 6 systems are AM\,CVns (see Sect.~\ref{subsec:AM_CVns}). This leaves 40 systems where the origin of the associated X-ray emission will be investigated in future work.

Based on $P_{\rm orb}$ measurements and photometric constraints on the donor SpT, we identified two systems
that were not observed in SDSS-V, USNO-A2.0 0525-17852977 and \textit{Gaia} DR3 2884625449040056704, as strong PB candidates (Appendix~\ref{sec:special_systems_Plots}), pending confirmation through follow-up observations.

To assess the evolutionary status of the PB candidates observed with SDSS-V, we modelled the WD contribution in the SDSS-V spectra using DA WD atmosphere models, deriving $T_{\rm eff}$ and $\langle\dot{M}\rangle$. The resulting values are consistent with those of known PBs (see Sect.~\ref{subsec:Teff_Mdot_Distributions}). We also measured the Balmer decrements, used as a diagnostic of the physical conditions in the accretion discs of non-magnetic CVs, which allows to discriminate PBs from short-period pre-bouncers. The lower mass-transfer rates in PBs lead to cooler, optically thinner discs, which is reflected in the Balmer decrements. Our results suggest that the currently known PB population may be biased toward the less evolved systems, with most of our candidates displaying steeper Balmer decrements compared to previously known PBs (see Sect.~\ref{subsec:BD_discussion}). Together with the modelling of the SED for a subset of 14 newly identified CVs, we inferred substellar donors in most cases (see Sect.~\ref{subsec:confirmation_PBs}). Follow-up observations (in particular aimed at obtaining measurements of $P_{\rm orb}$) are required for further confirmation. Considering the so far strong, yet inconclusive evidence of the PB status of these systems, follow-up observations could potentially increase the known PB population by $\sim$50\%. Overall, our study supports the hypothesis that a substantial fraction of PBs remains hidden in WD catalogues due to their intrinsic faintness, which is associated with low mass-accretion rates. 

Among the new CVs, we identified three systems that could represent PBs approaching detachment following the onset of WD magnetism (see Sect.~\ref{subsec:PBs_detached}). Although X-ray observations and time-resolved spectroscopy are required to understand the accretion state of these systems, this finding reinforces the possibility that the long-standing discrepancy between theoretical predictions and the observed number of PBs arises not only from an incomplete detection of these systems, but also from a fraction of PBs having already detached.

Finally, continued follow-up of the remaining candidates from \citet{Daniela_2026} is expected to yield a more robust assessment of the PB population. The 213 high-likelihood PB candidates are confined to 500\,pc, with $\sim$100 systems within 250\,pc. Given the eRASS:4 flux limit (2$\times$10$^{-14}$\,erg/cm$^2$/s; \citealt{Daniela_2026}) and the X-ray luminosity of the faintest PB (log(L$_{\rm x}$)=29.1 [erg/s]; \citealt{Daniela_2024}), our catalogue of high-likelihood PB candidates is expected to be highly complete out to $\sim$250\,pc. A complete, volume-limited sample within $\sim$250\,pc will enable a decisive test of whether the predicted PB fraction is accurate or overestimated, potentially pointing to missing physics in CV evolution models.

\section*{Data availability}
\label{sec:Data_availability}

Tables~\ref{table:Master}, \ref{table:WDs_parameters}, \ref{table:Balmer_Decrements}, and \ref{table:SED_Results} are only available in electronic form at the CDS via anonymous ftp to \href{http://cdsarc.u-strasbg.fr/}{cdsarc.u-strasbg.fr} (\href{ftp://130.79.128.5/}{130.79.128.5}) or via 
\url{http://cdsweb.u-strasbg.fr/cgi-bin/qcat?J/A+A/}.

\begin{acknowledgements}
We wish to thank the anonymous referee for constructive comments and suggestions that helped improve the clarity and quality
of this article. We are thankful to Jos\'e G. Fern\'andez-Trincado for input on this manuscript. SHD, JB and KP acknowledge  financial support from Deutsche Forschungsgemeinschaft (DFG) within research unit FOR 2990 under grant numbers  STE\,1068/6-2, Schw\,536/37-1 and Schw\,536/37-2. This project has received funding from the European Research Council (ERC) under the European Union’s Horizon 2020 research and innovation programme (Grant agreement No. 101020057). This work is based on data from eROSITA, the soft X-ray instrument aboard SRG, a joint Russian-German science mission supported by the Russian Space Agency (Roskosmos), in the interests of the Russian Academy of Sciences represented by its Space Research Institute (IKI), and the Deutsches Zentrum f\"{u}r Luft- und Raumfahrt (DLR). The SRG spacecraft was built by Lavochkin Association (NPOL) and its subcontractors, and is operated by NPOL with support from the Max Planck Institute for Extraterrestrial Physics (MPE). The development and construction of the eROSITA X-ray instrument was led by MPE, with contributions from the Dr. Karl Remeis Observatory Bamberg \& ECAP (FAU Erlangen-Nuernberg), the University of Hamburg Observatory, the Leibniz Institute for Astrophysics Potsdam (AIP), and the Institute for Astronomy and Astrophysics of the University of T\"{u}bingen, with the support of DLR and the Max Planck Society. The Argelander Institute for Astronomy of the University of Bonn and the Ludwig Maximilians Universit\"{a}t Munich also participated in the science preparation for eROSITA. The eROSITA data shown here were processed using the eSASS/NRTA software system developed by the German eROSITA consortium. 
This work has made use of data from the European Space Agency (ESA) mission {\it Gaia} (\url{https://www.cosmos.esa.int/gaia}), processed by the {\it Gaia} Data Processing and Analysis Consortium (DPAC, \url{https://www.cosmos.esa.int/web/gaia/dpac/consortium}). Funding for the DPAC has been provided by national institutions, in particular the institutions participating in the {\it Gaia} Multilateral Agreement. 
This publication makes use of VOSA, developed under the Spanish Virtual Observatory (\url{https://svo.cab.inta-csic.es}) project funded by MCIN/AEI/10.13039/501100011033/ through grant PID2020-112949GB-I00. VOSA has been partially updated by using funding from the European Union's Horizon 2020 Research and Innovation Programme, under Grant Agreement n$^\circ$ 776403 (EXOPLANETS-A).
Funding for the Sloan Digital Sky Survey V has been provided by the Alfred P. Sloan Foundation, the Heising-Simons Foundation, the National Science Foundation, and the Participating Institutions. SDSS acknowledges support and resources from the Center for High-Performance Computing at the University of Utah. SDSS telescopes are located at Apache Point Observatory, funded by the Astrophysical Research Consortium and operated by New Mexico State University, and at Las Campanas Observatory, operated by the Carnegie Institution for Science. The SDSS web site is \url{www.sdss.org}. SDSS is managed by the Astrophysical Research Consortium for the Participating Institutions of the SDSS Collaboration, including the Carnegie Institution for Science, Chilean National Time Allocation Committee (CNTAC) ratified researchers, Caltech, the Gotham Participation Group, Harvard University, Heidelberg University, The Flatiron Institute, The Johns Hopkins University, L'Ecole polytechnique f\'{e}d\'{e}rale de Lausanne (EPFL), Leibniz-Institut f\"{u}r Astrophysik Potsdam (AIP), Max-Planck-Institut f\"{u}r Astronomie (MPIA Heidelberg), Max-Planck-Institut f\"{u}r Extraterrestrische Physik (MPE), Nanjing University, National Astronomical Observatories of China (NAOC), New Mexico State University, The Ohio State University, Pennsylvania State University, Smithsonian Astrophysical Observatory, Space Telescope Science Institute (STScI), the Stellar Astrophysics Participation Group, Universidad Nacional Aut\'{o}noma de M\'{e}xico, University of Arizona, University of Colorado Boulder, University of Illinois at Urbana-Champaign, University of Toronto, University of Utah, University of Virginia, Yale University, and Yunnan University.
This work is based on data collected by the TESS mission and obtained from the MAST data archive at the Space Telescope Science Institute (STScI). Funding for the missions is provided by the NASA Explorer Program and the NASA Science Mission Directorate. STScI is operated by the Association of Universities for Research in Astronomy, Inc., under NASA contract NAS 5-26555.
The CSS survey is funded by the National Aeronautics and Space Administration under Grant No. NNG05GF22G issued through the Science Mission Directorate Near-Earth Objects Observations Program. The CRTS survey is supported by the U.S.~National Science Foundation under grants AST-0909182. 
This work has made use of data from the Asteroid Terrestrial-impact Last Alert System (ATLAS) project. The Asteroid Terrestrial-impact Last Alert System (ATLAS) project is primarily funded to search for near earth asteroids through NASA grants NN12AR55G, 80NSSC18K0284, and 80NSSC18K1575; byproducts of the NEO search include images and catalogues from the survey area. This work was partially funded by Kepler/K2 grant J1944/80NSSC19K0112 and HST GO-15889, and STFC grants ST/T000198/1 and ST/S006109/1. The ATLAS science products have been made possible through the contributions of the University of Hawaii Institute for Astronomy, the Queen’s University Belfast, the Space Telescope Science Institute, the South African Astronomical Observatory, and The Millennium Institute of Astrophysics (MAS), Chile. 
Based on observations obtained with the Samuel Oschin Telescope 48-inch and the 60-inch Telescope at the Palomar Observatory as part of the Zwicky Transient Facility project. ZTF is supported by the National Science Foundation under Grants No. AST-1440341 and AST-2034437 and a collaboration including current partners Caltech, IPAC, the Oskar Klein Center at Stockholm University, the University of Maryland, University of California, Berkeley , the University of Wisconsin at Milwaukee, University of Warwick, Ruhr University, Cornell University, Northwestern University and Drexel University. Operations are conducted by COO, IPAC, and UW. 
This research has made use of the NASA/IPAC Infrared Science Archive, which is funded by the National Aeronautics and Space Administration and operated by the California Institute of Technology. 

\end{acknowledgements}

%

\bibliography{BIB.bib}

\begin{appendix}





\onecolumn

\begin{table*}[h!]
\section{Sample description}
\label{sec:master_table}

\caption {Main properties of the sample of systems with evidence supporting a CV nature, as indicated by Balmer emission in the SDSS-V spectra, outburst activity in long-term light curves, or periodic variability detected in TESS light curves.}
\label{table:Master} 
\begin{threeparttable}
\begin{minipage}{0.60\textwidth}
\centering
\begin{tabular}{clll}
\hline\hline             
\# & Name & Unit & Description \\
\hline
1 & System &  & Object name most commonly used in the literature. \\
2 & GaiaDR3 &  & \textit{Gaia} ID from data release 3. \\
3 & ra\_DR3 & deg & \textit{Gaia}\,DR3 Right Ascension. \\
4 & dec\_DR3 & deg & \textit{Gaia}\,DR3 Declination. \\
5 & bp\_phot & mag &  \textit{Gaia} $BP$ mean magnitude. \\
6 & rp\_phot & mag & \textit{Gaia} $RP$ mean magnitude. \\
7 & g\_phot & mag & \textit{Gaia} $G$ mean magnitude. \\
8 & Distance & pc & Distance to the object (1). \\
9 & lo\_e\_Distance & pc & Lower uncertainty in Distance (1). \\
10 & up\_e\_Distance & pc & Upper uncertainty in Distance (1). \\
11 & Lx & $10^{30}\,$erg/s & eRASS:4 X-ray luminosity in the 0.2-2.3 keV band.\\
12 & log\_Fx/Fopt &  & Logarithm of the X-ray-to-optical flux ratio. \\
13 & SDSS-V\_flag &  & Flag indicating whether the object has been observed in SDSS-V. \\
14 & Outburst\_flag &  & Flag indicating whether an outburst is detected in long-term light curves\tablefootmark{a}. \\
15 & Period\_flag &  & Flag indicating whether a period has been detected in TESS light curves. \\
16 & PB\_Score\_reduced &  & Score from the multi-wavelength reduced PB scorecard (expressed as a percentage)\tablefootmark{b}.  \\
17 & PB\_Status &  & Evolutionary status of the system. \\
18 & PB\_Status\_Reference &  & Reference for the evolutionary status of the system. \\
\hline
\end{tabular}
    \tablefoot{The table lists coordinates and photometry from \textit{Gaia}, distances, and X-ray properties from eROSITA. Flags indicate SDSS-V spectroscopic coverage, identification of outburst activity, and period detection (Y: yes; N: no). We also include the score from the multi-wavelength reduced PB scorecard, the PB status of the system (PB: period-bouncer; PBc: period-bouncer candidate) and the corresponding reference. The full Table is available at the CDS. \tablefoottext{a}{The new outbursting systems were identified in CRTS (GALEX\,J040338.2-104945), ATLAS (\textit{Gaia} DR3 2902110913736006144, \textit{Gaia} DR3 2893005857946801152, \textit{Gaia} DR3 5561549440740425216, \textit{Gaia} DR3 6188581062130914560, and GALEX\,J224117.4-675915), CRTS and ATLAS (\textit{Gaia} DR3 3985446320286121472), and CRTS, ASAS-SN, and ATLAS (USNO-A2.0 0525-17852977).}\tablefoottext{b}{The multi-wavelength PB scorecard is defined by \citet{Daniela_2024}.}}
    \tablebib{(1) \citet{Bailer-Jones}.}
\end{minipage}
\end{threeparttable}
\end{table*}

\begin{table*}[h!]
\centering
\section{AM\,CVn systems}
\label{sec:AM_CVns_table}
\begin{threeparttable}
\begin{minipage}{0.60\textwidth}
    \caption{AM\,CVn systems identified in our sample of PB candidates.}
    \label{table:AM_CVns}
    \centering
    \begin{tabular}{l c c c}
        \hline\hline
        System & SDSS-V\_flag & Outburst\_flag & Reference \\
        \hline
       SDSS\,J090344.24-013326.1 & Y & N & (1) \\ 
       GP\,Com & Y & N & (2) \\
       GALEX\,J085934.2-111054  & Y & N & This work \\
       GALEX\,J044842.7-442330  & Y & N & This work \\
       \textit{Gaia} DR3 5044321750645546112  & Y & N & This work \\  
       GALEX\,J112611.6+154926\tablefootmark{a}  & Y & N & This work \\
       3XMM\,J035517.9-822609 & N & Y & (3) \\
       ASASSN -14mv & N & Y & (4) \\
       \textit{Gaia} DR3 3124112584945910912 & N & Y & (5) \\
        \hline
    \end{tabular}
    \tablefoot{Columns include the object name most commonly used in the literature, whether the object was observed by SDSS-V, whether an outburst was detected in the long-term light curves, and the literature reference for the classification of the system. Flags are indicated as Y (yes) or N (no). \tablefoottext{a}{The SDSS-V spectrum of GALEX\,J112611.6+154926 is noisy, and its classification as an AM\,CVn is therefore tentative.}}
    \tablebib{(1) \citet{Inight_2023}, (2) \citet{Nather_1981}, (3) \citet{Kato_2015b}, \newline (4) \citet{Ramsay_2018}, (5) \citet{Maehara_2023}.}
\end{minipage}
\end{threeparttable}
\end{table*}

\begin{table*}[h!]
\section{WDs parameters from spectroscopic modelling}
\label{sec:WDs_parameters}

\caption {Results from the spectroscopic modelling of the WD component. 
\newline}
\label{table:WDs_parameters} 
\begin{threeparttable}
\begin{minipage}{0.60\textwidth}
\centering
\begin{tabular}{clll}
\hline\hline             
\# & Name & Unit & Description \\
\hline
1 & System &  & Object name most commonly used in the literature. \\
2 & GaiaDR3 &  & \textit{Gaia} ID from data release 3. \\
3 & MJD &  & Modified Julian Date of the SDSS-V spectroscopic observation. \\
4 & Teff\_WD & K & Effective temperature of the white dwarf. \\
5 & lo\_e\_Teff\_WD & K & Lower uncertainty in Teff\_WD. \\
6 & hi\_e\_T\_eff\_WD & K & Upper uncertainty in Teff\_WD. \\
7 & logg\_WD &  & Surface gravity of the white dwarf ($\log(g)$ in cgs units). \\
8 & lo\_e\_logg\_WD &  & Lower uncertainty in logg\_WD, in dex. \\
9 & hi\_e\_logg\_WD &  & Upper uncertainty in logg\_WD, in dex. \\
10 & K\_WD &  & Dilution factor of the white dwarf, in units of $10^{-25}$. \\
11 & lo\_e\_K\_WD &  & Lower uncertainty in K\_WD, in units of $10^{-25}$. \\
12 & hi\_e\_K\_WD &  & Upper uncertainty in K\_WD, in units of $10^{-25}$. \\
13 & R\_WD & R$_{\odot}$ & Radius of the white dwarf. \\
14 & lo\_e\_R\_WD & R$_{\odot}$ & Lower uncertainty in R\_WD. \\
15 & hi\_e\_R\_WD & R$_{\odot}$ & Upper uncertainty in R\_WD. \\
16 & M\_WD & M$_{\odot}$ & Mass of the white dwarf. \\
17 & lo\_e\_M\_WD & M$_{\odot}$ & Lower uncertainty in M\_WD. \\
18 & hi\_e\_M\_WD & M$_{\odot}$ & Upper uncertainty in M\_WD. \\
19 & M\_dot & $10^{-10}\,M_{\odot}\,\mathrm{yr^{-1}}$ & Secular mean mass accretion rate onto the white dwarf. \\
20 & lo\_e\_M\_dot & $10^{-10}\,M_{\odot}\,\mathrm{yr^{-1}}$ & Lower uncertainty in M\_dot. \\
21 & hi\_e\_M\_dot & $10^{-10}\,M_{\odot}\,\mathrm{yr^{-1}}$ & Upper uncertainty in M\_dot. \\
22 & GALEX\_flag &  & Flag indicating whether GALEX photometry was included in the spectroscopic fit. \\
\hline
\end{tabular}
    \tablefoot{The table presents the best-fit atmospheric parameters, together with derived physical parameters of the WD. Uncertainties are reported as asymmetric lower and upper bounds. A flag indicates if GALEX photometry was used in the spectroscopic fit (Y: yes; N: no). The full Table is available at the CDS.}
\end{minipage}
\end{threeparttable}
\end{table*}

\begin{table*}[h!]
\section{Balmer emission line measurements and decrements}
\label{sec:balmer_decrements_results}

\caption {Results from the Balmer line analysis of the SDSS-V spectra. 
\newline}
\label{table:Balmer_Decrements} 
\begin{threeparttable}
\begin{minipage}{0.60\textwidth}
\centering
\begin{tabular}{clll}
\hline\hline             
\# & Name & Unit & Description \\
\hline
1 & System &  & Object name most commonly used in the literature. \\
2 & GaiaDR3 &  & \textit{Gaia} ID from data release 3. \\
3 & MJD &  & Modified Julian Date of the SDSS spectroscopic observation. \\
4 & F\_Halpha & $10^{-17}\,$erg/cm$^{2}$/s & $\text{H}\alpha$ line flux. \\
5 & e\_F\_Halpha & $10^{-17}\,$erg/cm$^{2}$/s & Uncertainty in F\_Halpha. \\
6 & F\_Hbeta & $10^{-17}\,$erg/cm$^{2}$/s & $\text{H}\beta$ line flux. \\
7 & e\_F\_Hbeta & $10^{-17}\,$erg/cm$^{2}$/s & Uncertainty in F\_Hbeta. \\
8 & F\_Hgamma & $10^{-17}\,$erg/cm$^{2}$/s & $\text{H}\gamma$ line flux. \\
9 & e\_F\_Hgamma & $10^{-17}\,$erg/cm$^{2}$/s & Uncertainty in F\_Hgamma. \\
10 & L\_Halpha & $10^{27}\,$erg/s & $\text{H}\alpha$ line luminosity. \\
11 & lo\_e\_L\_Halpha & $10^{27}\,$erg/s & Lower uncertainty in L\_Halpha. \\
12 & up\_e\_L\_Halpha & $10^{27}\,$erg/s & Upper uncertainty in L\_Halpha. \\
13 & L\_Hbeta & $10^{27}\,$erg/s & $\text{H}\beta$ line luminosity. \\
14 & lo\_e\_L\_Hbeta & $10^{27}\,$erg/s & Lower uncertainty in L\_Hbeta. \\
15 & up\_e\_L\_Hbeta & $10^{27}\,$erg/s & Upper uncertainty in L\_Hbeta. \\
16 & L\_Hgamma & $10^{27}\,$erg/s & $\text{H}\gamma$ line luminosity. \\
17 & lo\_e\_L\_Hgamma & $10^{27}\,$erg/s & Lower uncertainty in L\_Hgamma. \\
18 & up\_e\_L\_Hgamma & $10^{27}\,$erg/s & Upper uncertainty in L\_Hgamma. \\
19 & Halpha/Hbeta &  & Luminosity ratio between H$\alpha$ and H$\beta$ lines. \\
20 & e\_Halpha/Hbeta &  & Uncertainty in Halpha/Hbeta. \\
21 & Hgamma/Hbeta &  & Luminosity ratio between H$\gamma$ and H$\beta$ lines. \\
22 & e\_Hgamma/Hbeta &  & Uncertainty in Hgamma/Hbeta. \\
23 & P\_PB &  & PB probability (expressed as a percentage)\tablefootmark{a}. \\
\hline
\end{tabular}
    \tablefoot{The table lists Balmer line fluxes, Balmer line luminosities, Balmer decrements, and PB probabilities. The full Table is available at the CDS. \tablefoottext{a}{PB probabilities were calculated using the logistic regression model developed in \citet{Hernandez-Diaz_2026}.}}
\end{minipage}
\end{threeparttable}
\end{table*}

\begin{table*}[h!]
\section{Results from the search for orbital periods}
\label{sec:orbital_periods_results}
\begin{threeparttable}
\begin{minipage}{1.0\textwidth}

\caption{Results from the search for P$_{\text{orb}}$ using TESS two-minute cadence light curves.} 
\label{table:orbital_periods}
\centering
\begin{tabular}{l l l c c c c c}
\hline\hline
System & TIC &  Sectors & Period [min] & S/N$_{\text{PSD}}$ &  Reliable  & Interpretation & Reference \\
\hline
GALEX\,J125751.4-283015 &  999085360 & 37, 64 & $82.17\pm0.63$ & 0.01 & Y & $P_{\text{orb}}$ & (1) \\
BW\,Scl &  183676876 & 29 & $78.2239\pm0.0012$ & 0.07 & Y & $P_{\text{orb}}$ & (2) \\
EU\,UMa &  138848507 & 49 & $90.1884\pm0.0050$ & 0.09 & Y & $P_{\text{orb}}$ & (3) \\
 EF\,Eri & 137884253 & 31 & $81.010\pm0.023$ & 0.004 & Y & $P_{\text{orb}}$ & (4) \\
USNO-A2.0 0525-17852977 & 379593642 & 65 & $90.5724\pm0.0064$ & 0.01 & Y & $P_{\text{orb}}$ & - \\
\textit{Gaia} DR3 2884625449040056704 & 705387284 & 87, 98 & $82.3059\pm0.0051$ & 0.004 & Y & $P_{\text{orb}}$ & - \\
PNV\,J05580574-0011155 & 39580395  & 87 & $86.073\pm0.046$ & 0.002 & Y? & $P_{\text{orb}}$ & - \\
GALEX\,J022318.3-431701 & 631854368  & 97 & $39.539\pm0.014$ & 0.001 & Y? & $\frac{1}{2}P_{\text{orb}}$  & - \\

\hline
\end{tabular}
\tablefoot{The table lists the sectors used to derive the reported periods, the detected periodicities, their corresponding $\mathrm{S/N}_{\mathrm{PSD}}$ values, and a flag indicating whether each detection is considered reliable (Y: reliable; Y?: tentative). We also provide the physical interpretation of the periodicities detected and the references to prior measurements in the literature.}
\tablebib{(1) \citet{Daniela_2026}, (2) \citet{Augusteijn_1997}, (3) \citet{Howell_1995}, (4) \citet{Bailey_1982}.}
\end{minipage}
\end{threeparttable}
\end{table*}

\begin{table*}[h!]
\centering
\section{Results from the SED analysis}
\label{sec:SED_Results}
\begin{minipage}{0.63\textwidth}
\caption{Results from the SED fitting analysis.}
\label{table:SED_Results} 
\centering
\begin{tabular}{clll}
\hline\hline             
\# & Name & Unit & Description \\
\hline
1 & System &  & Object name most commonly used in the literature. \\
2 & GaiaDR3 &  & \textit{Gaia} ID from data release 3. \\
3 & Teff\_WD & K & Effective temperature of the white dwarf. \\
4 & e\_Teff\_WD & K & Uncertainty in Teff\_WD\tablefootmark{a}. \\
5 & Teff\_donor & K & Effective temperature of the donor. \\
6 & e\_Teff\_donor & K & Uncertainty in Teff\_donor\tablefootmark{a}. \\
7 & SpT\_donor &  & Spectral type of the donor\tablefootmark{b}. \\
\hline
\end{tabular}
\tablefoot{The table lists the best-fitting $T_{\rm eff}$ values for the WD and donor, together with their associated uncertainties, as well as the inferred SpT of the donor. The full Table is available at the CDS.
\tablefoottext{a}{The uncertainties returned by the VOSA binary fitting routine (\citealt{Bayo_2008}) are defined as half the adopted model-grid spacing and therefore correspond to numerical resolution limits rather than to statistical confidence intervals, likely underestimating the true statistical uncertainties.}
\tablefoottext{b}{The spectral type of the donor is inferred from the semi-empirical evolutionary tracks of \citet{Knigge_2011}.}}
\end{minipage}
\end{table*}

\begin{table*}[h!]
\centering
\section{Results from the Zeeman splitting analysis}
\label{sec:Zeeman_splitting_results}
\begin{minipage}{0.5\textwidth}
    \caption{Mean photospheric magnetic field strength values derived from the Zeeman splitting analysis in SDSS-V spectra.}
    \label{table:magnetic fields}
    \centering
    \renewcommand{\arraystretch}{1.4}
    \begin{tabular}{l c c c}
        \hline\hline
        System & MJD & Fitted line & $\langle B \rangle$\,[MG] \\
        \hline
        GALEX J033833.9-332802 & 60612 & H$\alpha$ & $9.6^{+0.1}_{-0.5}$ \\
        CP\,Tuc & 60576 & H$\alpha$ & $11.8^{+0.3}_{-0.4}$ \\
        CP\,Tuc & 60577 & H$\alpha$ & $12.4^{+0.2}_{-0.9}$ \\ 
        V379\,Vir & 60411 & H$\beta$ &  $4.7^{+0.1}_{-0.2}$ \\
        2QZ\,J130249.7+014919\tablefootmark{a} & 60382 &  H$\beta$ &  $9.8^{+0.7}_{-0.2}$ \\
        \multicolumn{3}{c}{} & $15.4^{+0.1}_{-0.8}$ \\
        \hline
    \end{tabular}
    \tablefoot{For each spectrum, we list the object's common name, the MJD at which the spectrum was observed, the Balmer line used in the fit, and the best-fitting $\langle B \rangle$ value. The derived $\langle B \rangle$ values represent mean photospheric magnetic field strengths. For a pure dipole field, the field strength can vary by a factor of two between the magnetic pole and equator (see e.g. \citealt{Martin_1984}).
    \tablefoottext{a}{For 2QZ\,J130249.7+014919, two distinct $\langle B \rangle$ values are associated with statistically comparable minima in the $\chi^{2}(B_{k})$ curve ($\Delta\chi^2 \leq 4$). Both solutions are therefore reported.}}
\end{minipage}
\end{table*}

\begin{figure}

    \centering
    
    \includegraphics[width=0.49\textwidth]{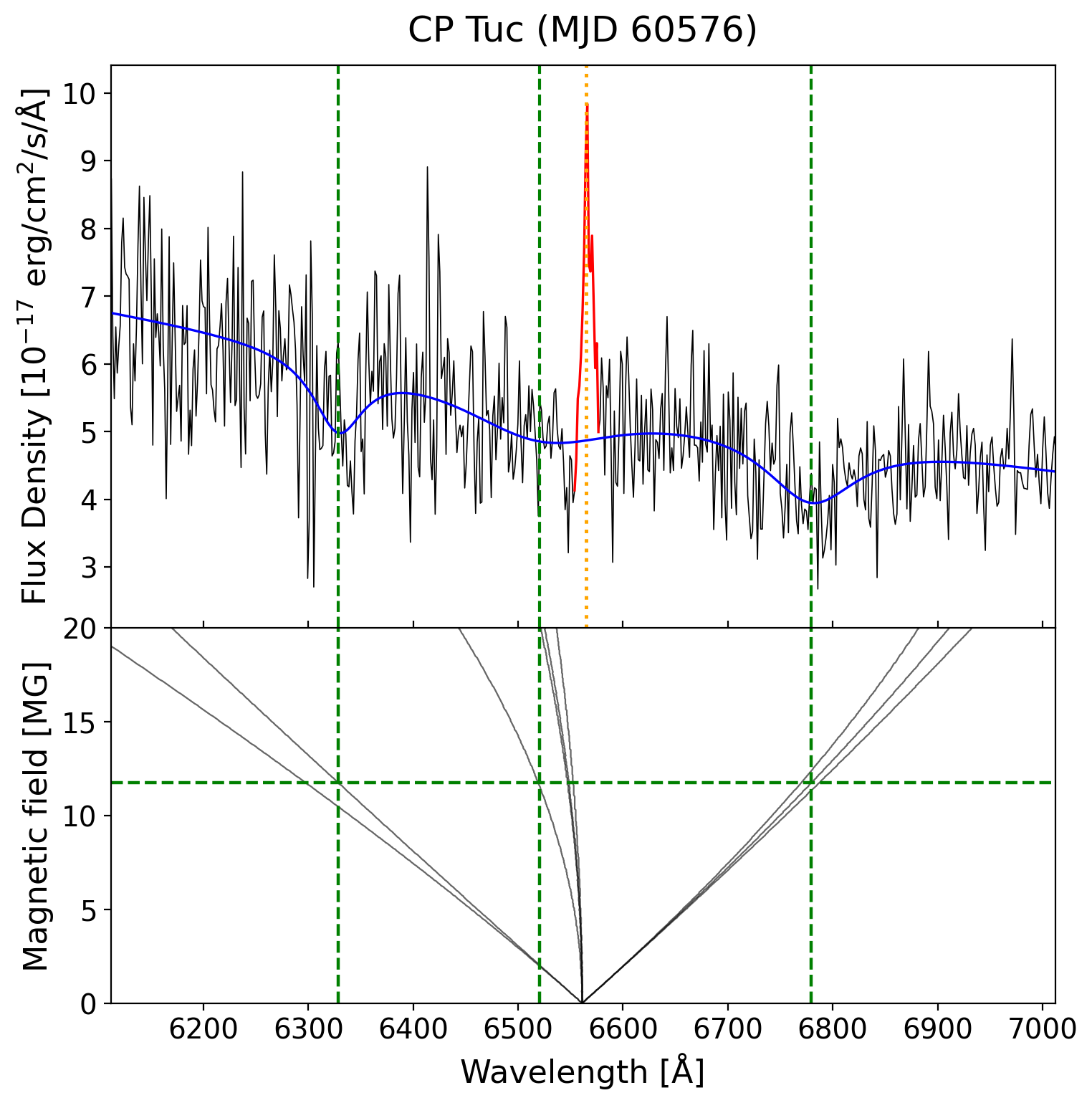}
    \includegraphics[width=0.49\textwidth]{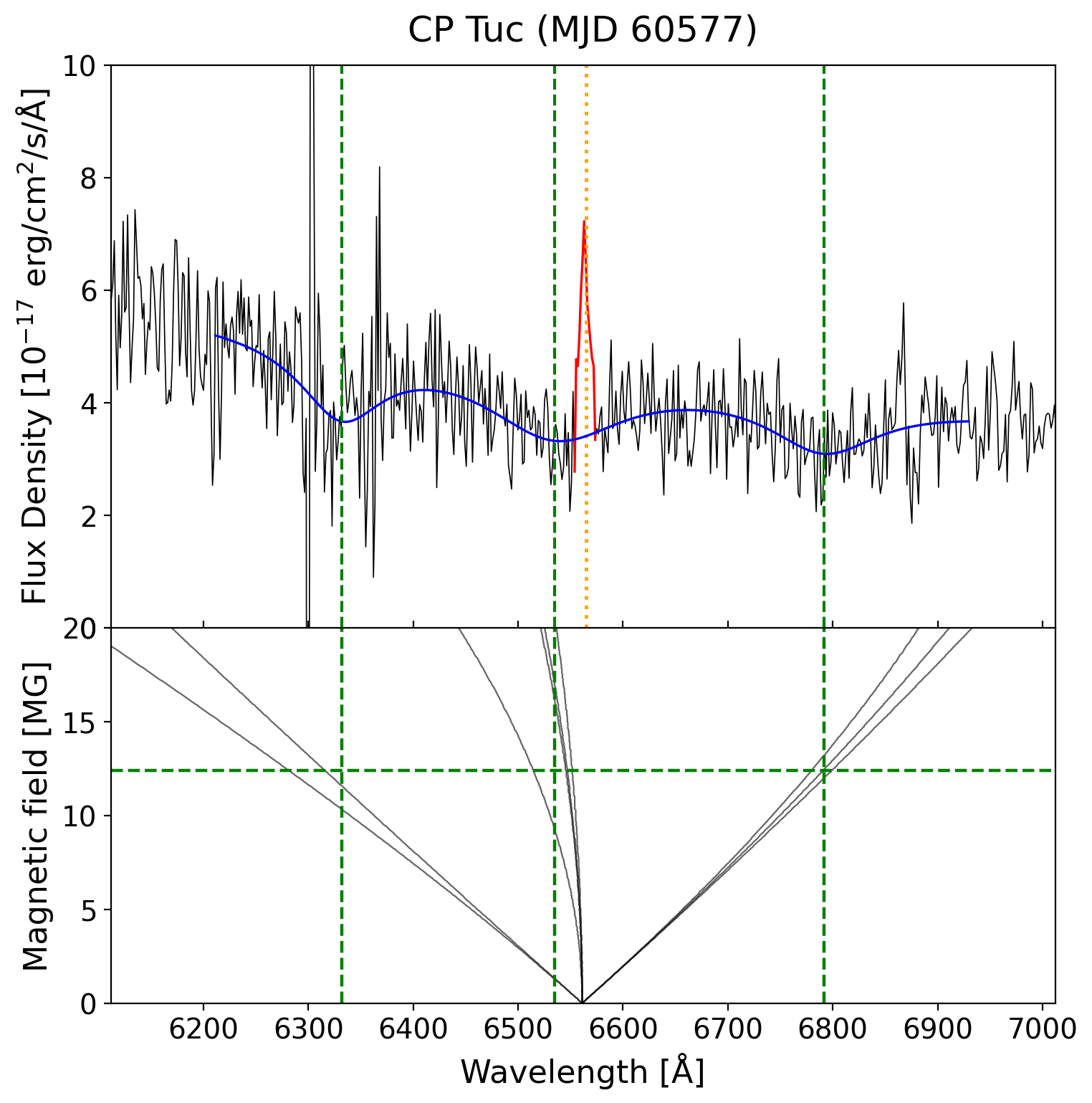}

    \vspace{0.4cm}

    \includegraphics[width=0.49\textwidth]{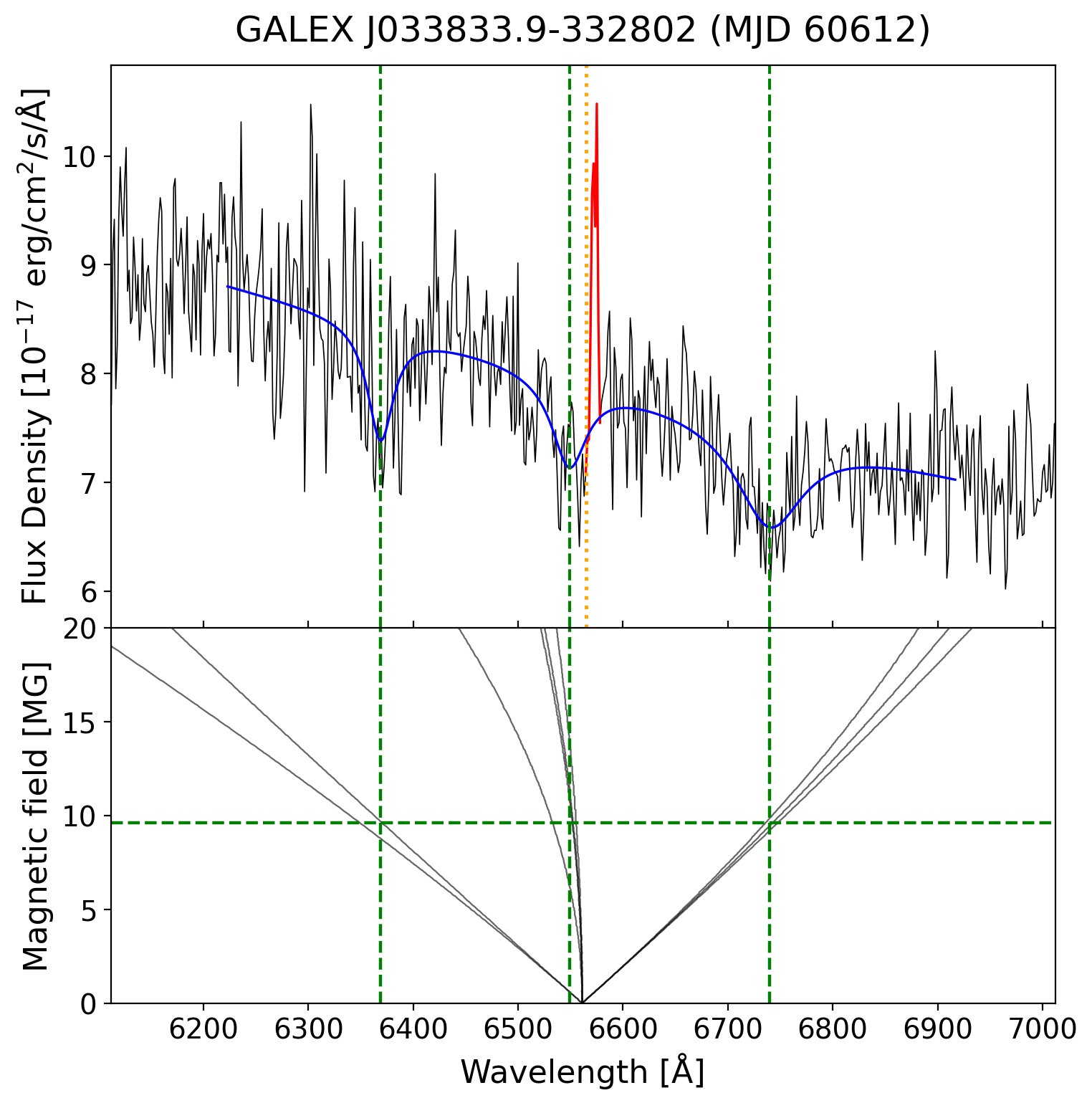}
    \includegraphics[width=0.49\textwidth]{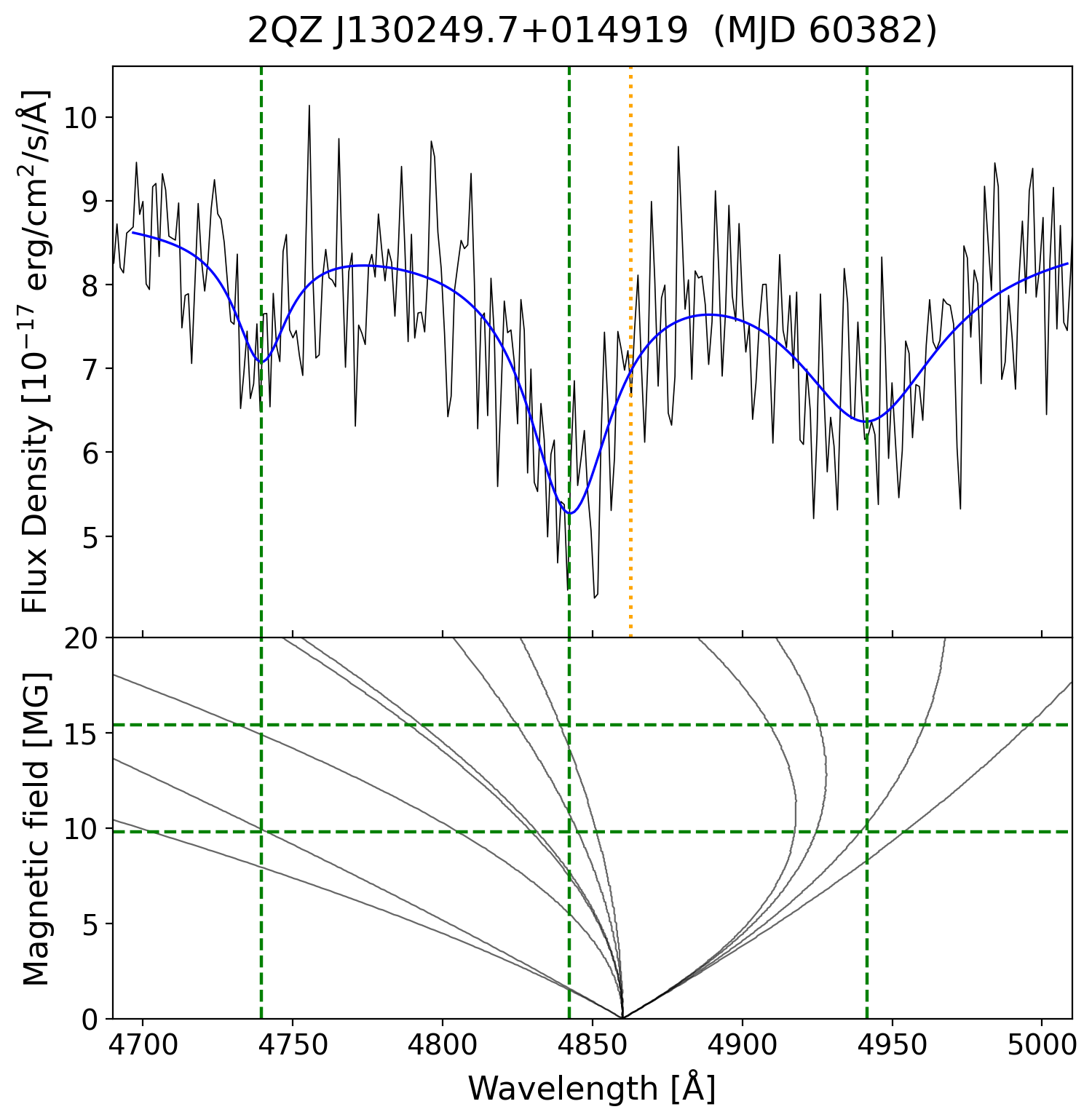}

    \caption{SDSS-V spectra of CP\,Tuc (top panels), GALEX J033833.9$-$332802 (bottom left), and 2QZ J130249.7+014919 (bottom right) in the vicinity of the H$\alpha$ line (CP\,Tuc and GALEX\,J033833.9$-$332802) and the H$\beta$ line (2QZ\,J130249.7+014919), together with the Zeeman-splitting fits used to estimate the mean photospheric magnetic field strength of the WD. The corresponding analysis for V379\,Vir is shown in Fig.~\ref{Fig.Zeeman_Splitting} in Sect.~\ref{subsec:Zeeman_Splitting}. The layout and elements shown in these figures are the same as described in Fig.~\ref{Fig.Zeeman_Splitting}.}
    
    \label{fig:Zeeman_Splitting_Plots}
\end{figure}

\onecolumn
\section{USNO-A2.0 0525-17852977 and \textit{Gaia} DR3 2884625449040056704}
\label{sec:special_systems_Plots}

\begin{figure*}[h!]
    \centering
    \includegraphics[width=0.49\textwidth]{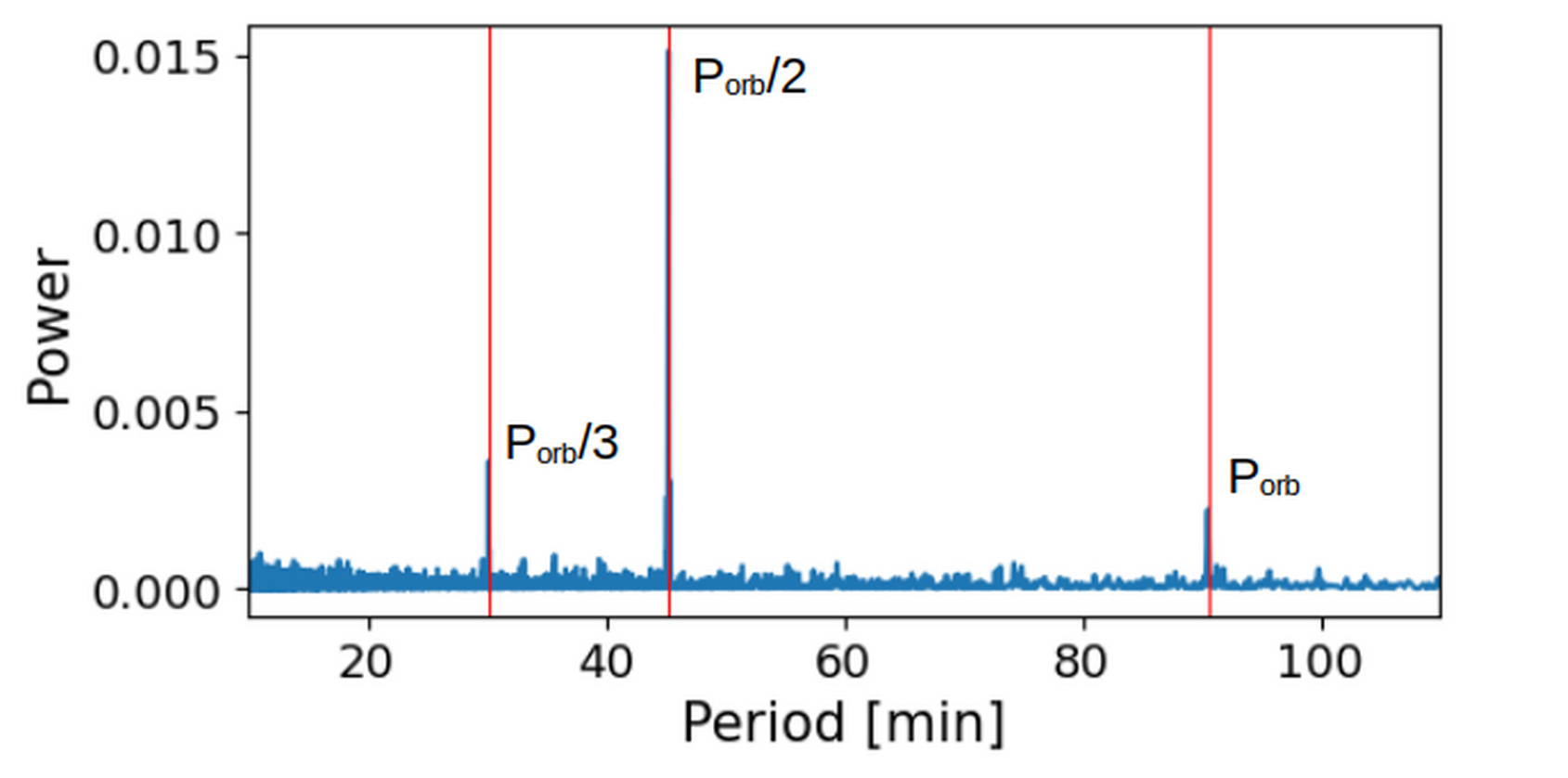}
    \includegraphics[width=0.49\textwidth]{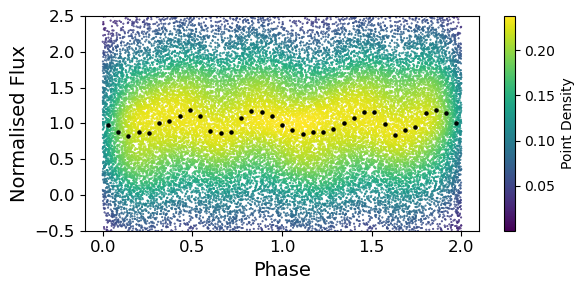}
    \includegraphics[width=0.54\textwidth]{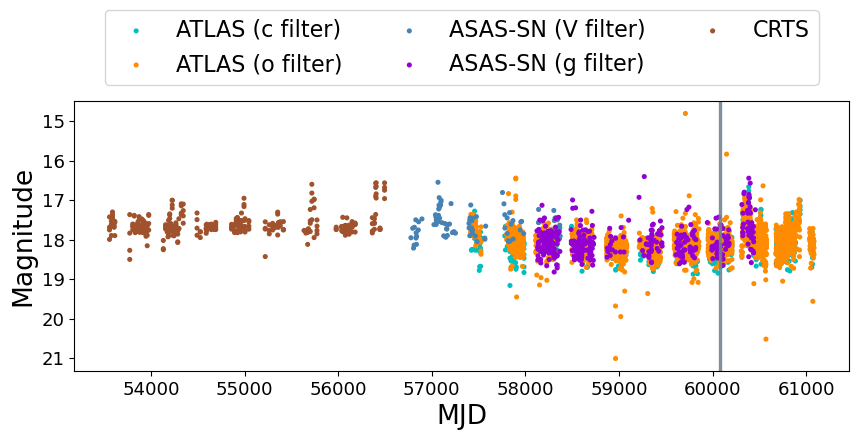}\hfill
    \includegraphics[width=0.44\textwidth]{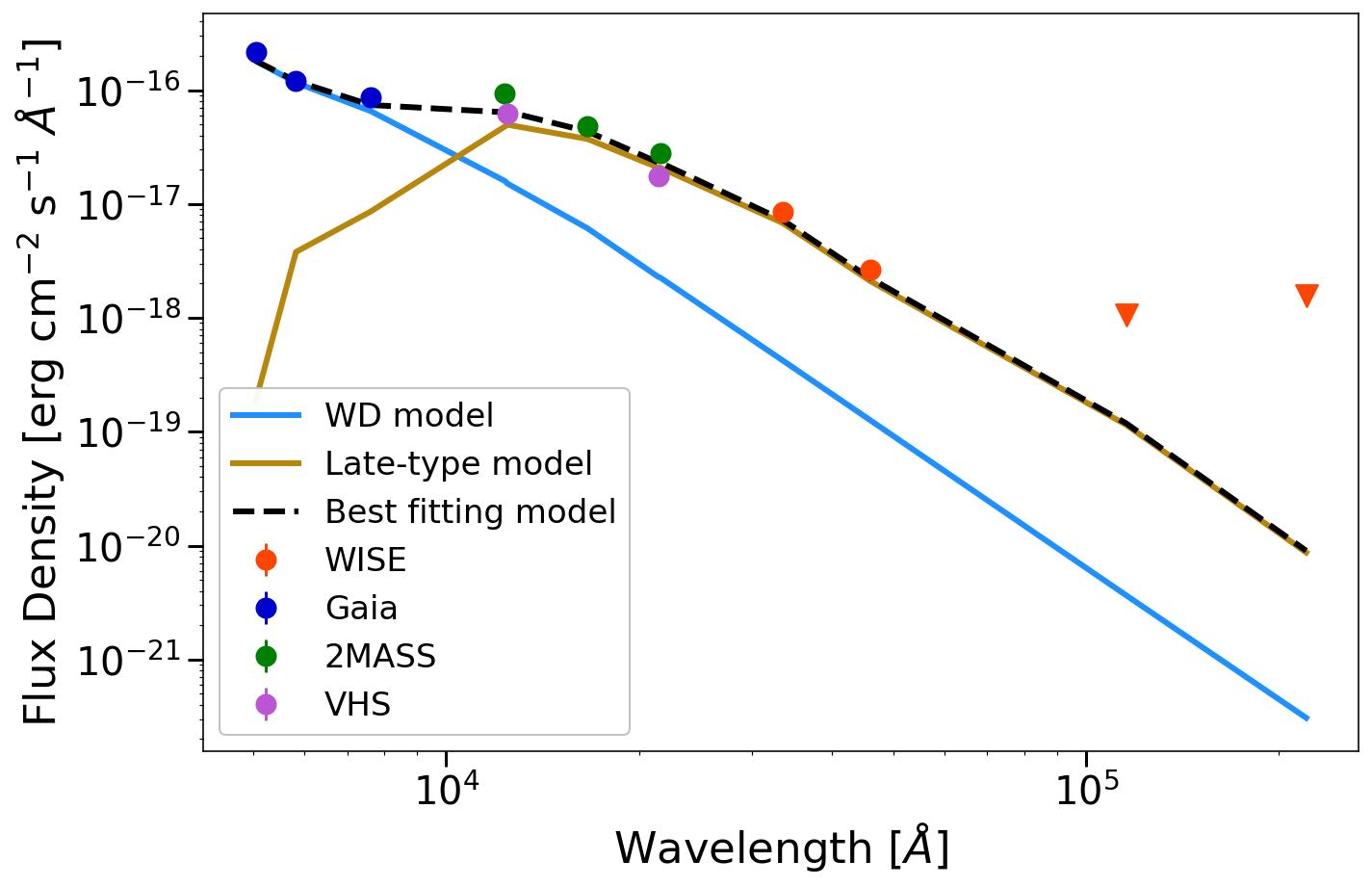}

    \caption{PB candidate USNO-A2.0 0525-17852977. Upper left panel: Lomb–Scargle periodogram of the TESS light curve (sector\,65). The red vertical lines mark the positions of the detected periodicities, interpreted as the orbital period, $P_{\rm orb}$, and its second and third harmonics.
    Upper right panel: Phase folded TESS light curve (sector\,65) for a period of 90.5724\,min, displaying a double-humped morphology. Black points indicate the phase-folded light curve binned into 18 equal phase bins.
    Bottom left panel: Long-term light curve showing outburst activity at multiple epochs. The vertical gray band indicates the observation time interval covered by the TESS sector\,65 light curve. 
    Bottom right panel: Spectral energy distribution fitted using a hydrogen WD model (T$_{\rm eff}$ = 10000\,K) and a late-type star model (T$_{\rm eff}$ = 1850\,K). The triangles indicate upper limits. The donor is inferred to be of SpT L4 or later.}
    \label{fig:Gaia DR3 6217118886429978112}
\end{figure*}

\begin{figure*}[h!]
    \centering
    \includegraphics[width=0.49\textwidth]{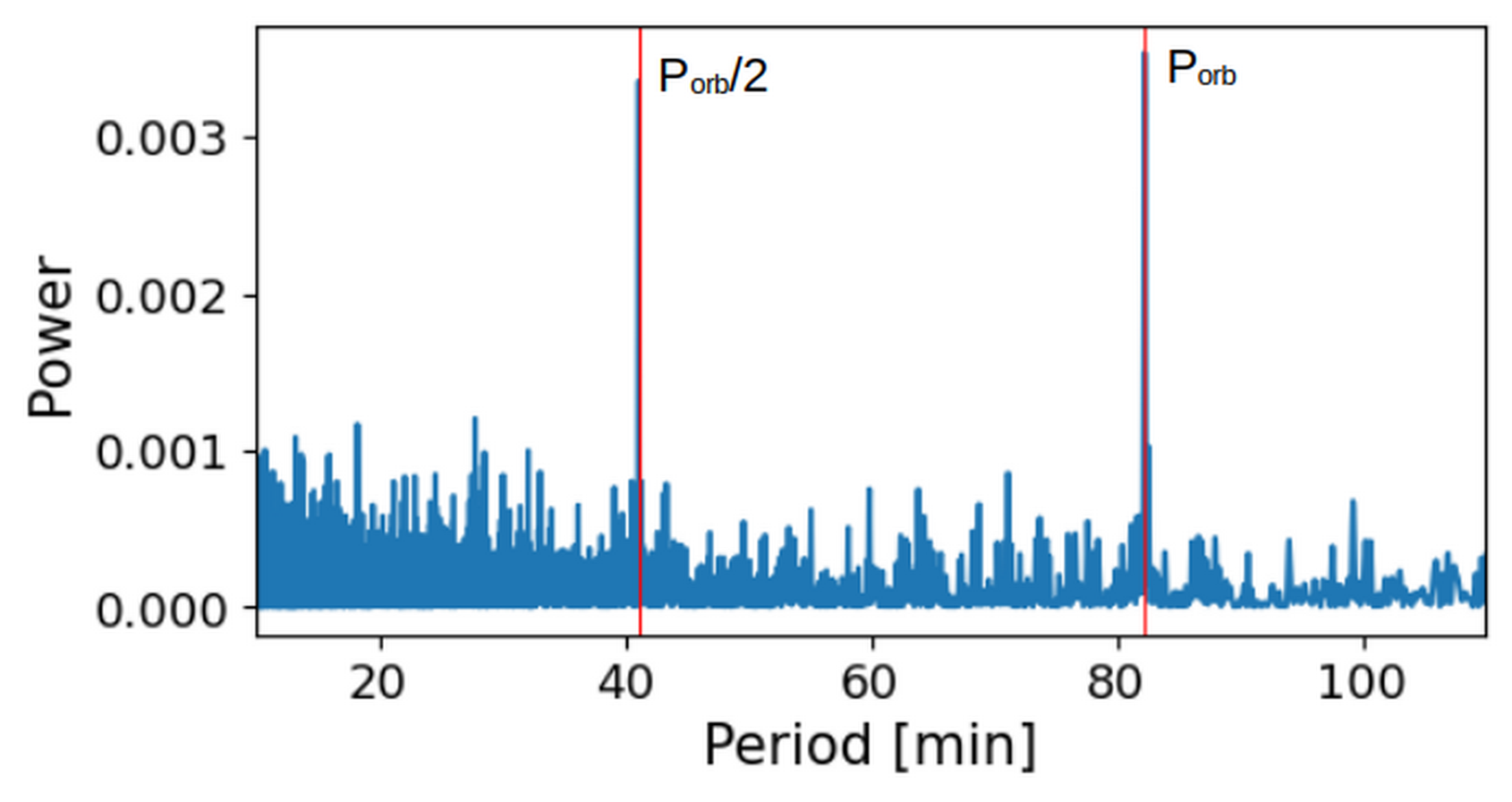}
    \includegraphics[width=0.49\textwidth]{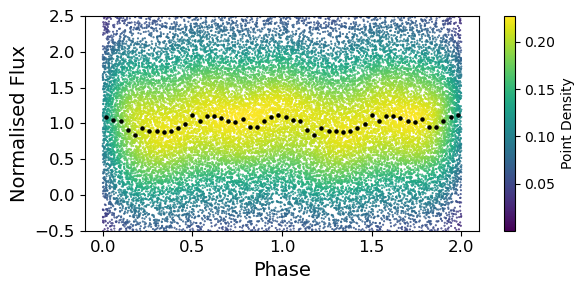}
    \includegraphics[width=0.54\textwidth]{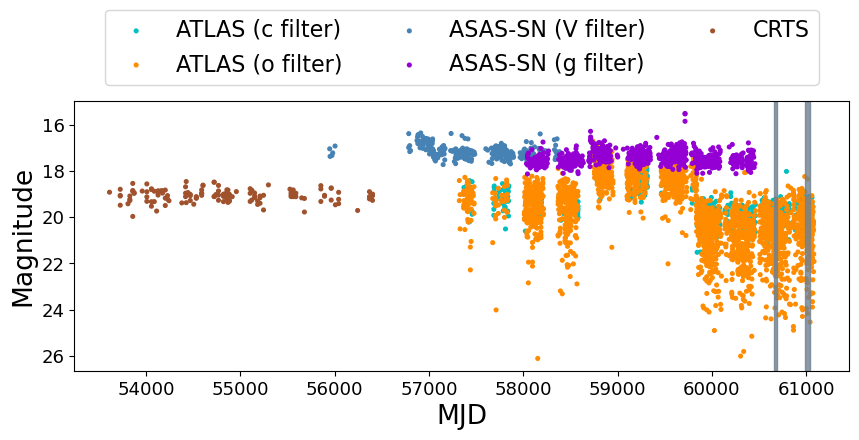}\hfill
    \includegraphics[width=0.44\textwidth]{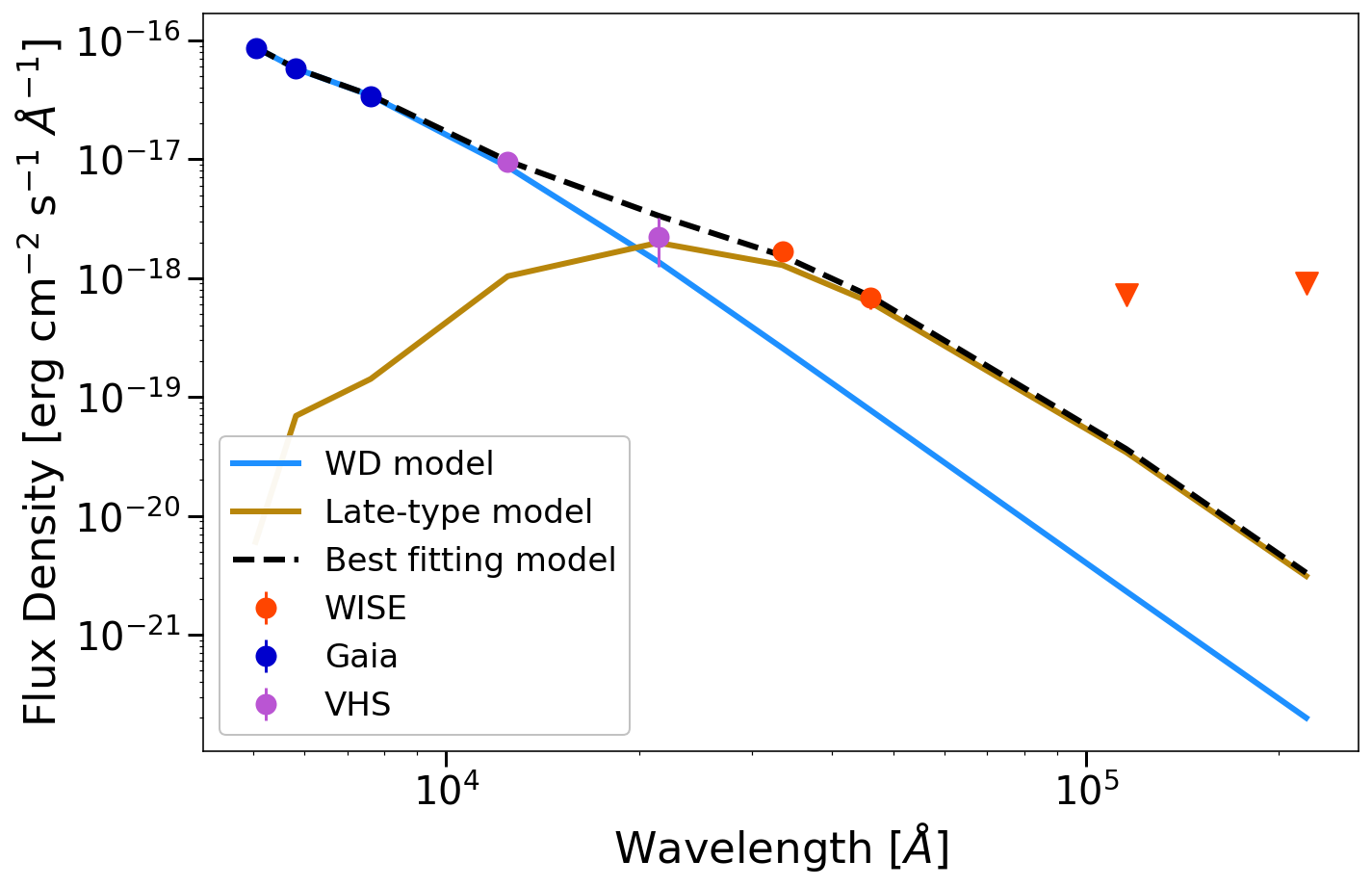}    
    \caption{PB candidate \textit{Gaia} DR3 2884625449040056704. 
    Upper left panel: Lomb–Scargle periodogram of the TESS light curve (sector\,87). The red vertical lines mark the positions of the detected periodicities, interpreted as the orbital period, $P_{\rm orb}$, and its second harmonic.
    Upper right panel: Phase folded TESS light curve (sector\,87) for a period of 82.5258\,min, displaying a double-humped morphology. Black points indicate the phase-folded light curve binned into 25 equal phase bins.
    Bottom left panel: Long-term light curve showing the transition between high and low states characteristic of mCVs. The vertical gray band indicates the observation time intervals covered by the TESS sector\,87 and sector\,98 light curves. 
    Bottom right panel: Spectral energy distribution fitted using a hydrogen WD model (T$_{\rm eff}$ = 9000\,K) and a late-type star model (T$_{\rm eff}$ = 1500\,K). The triangles indicate upper limits. The donor is inferred to be of SpT T.}
    \label{fig:Gaia DR3 884625449040056704}
\end{figure*}

\begin{figure}
\section{Optical spectra of candidates for PBs approaching detachment}
\label{sec:spectra_semi-detached}

    \centering
    
    \includegraphics[width=0.49\textwidth]{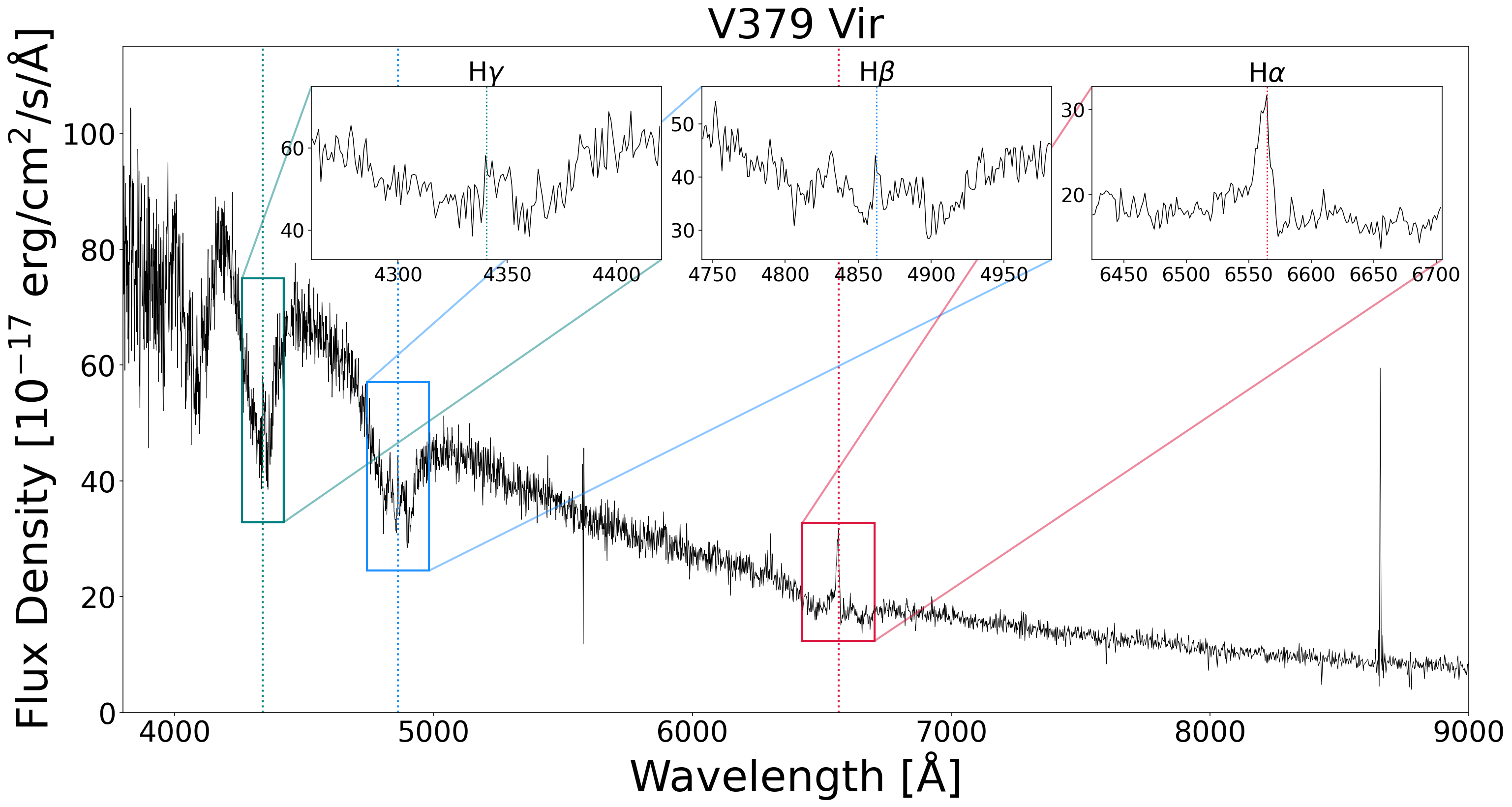}
    \includegraphics[width=0.49\textwidth]{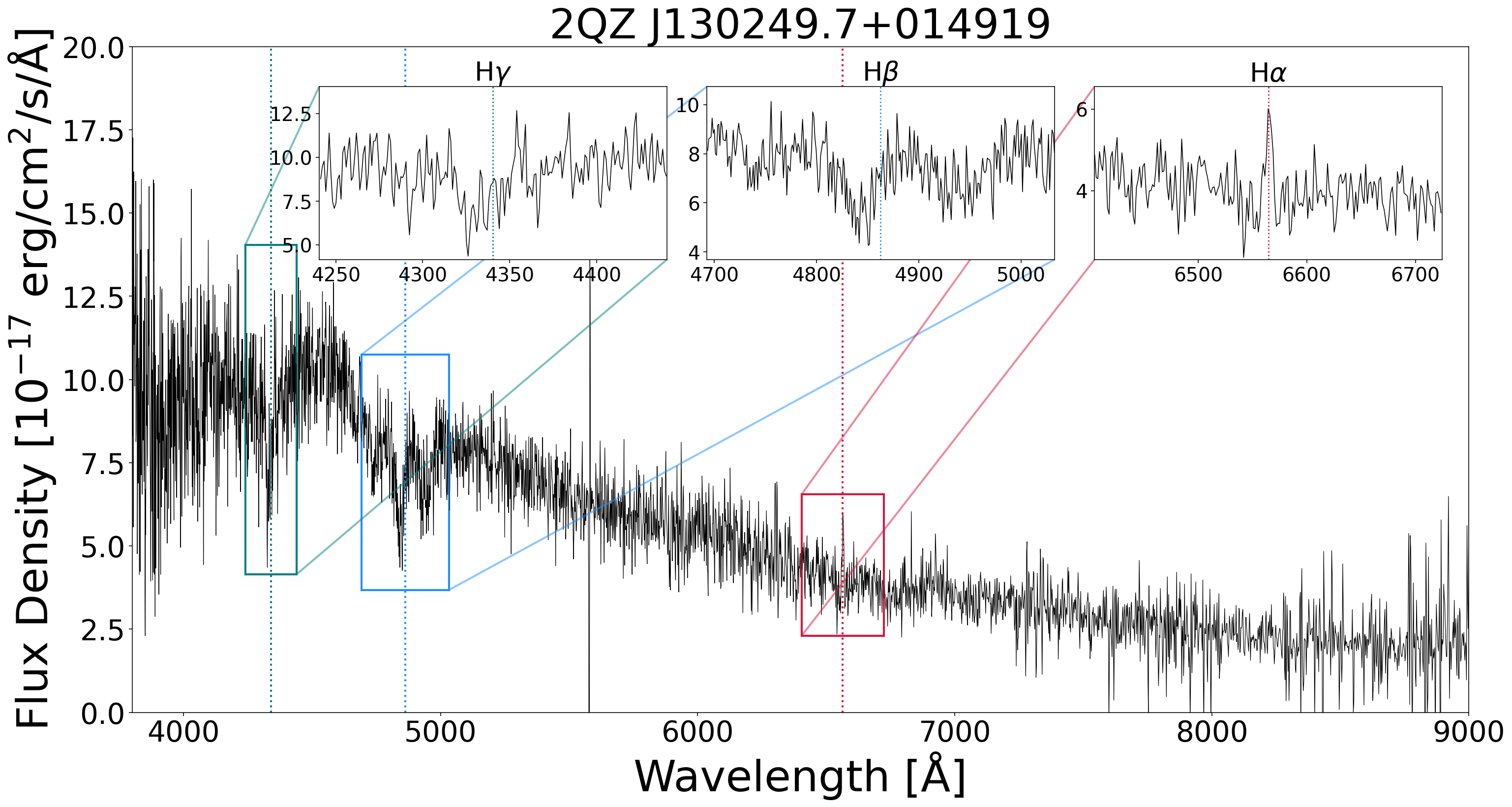}
    
    \vspace{0.4cm}

    \includegraphics[width=0.49\textwidth]{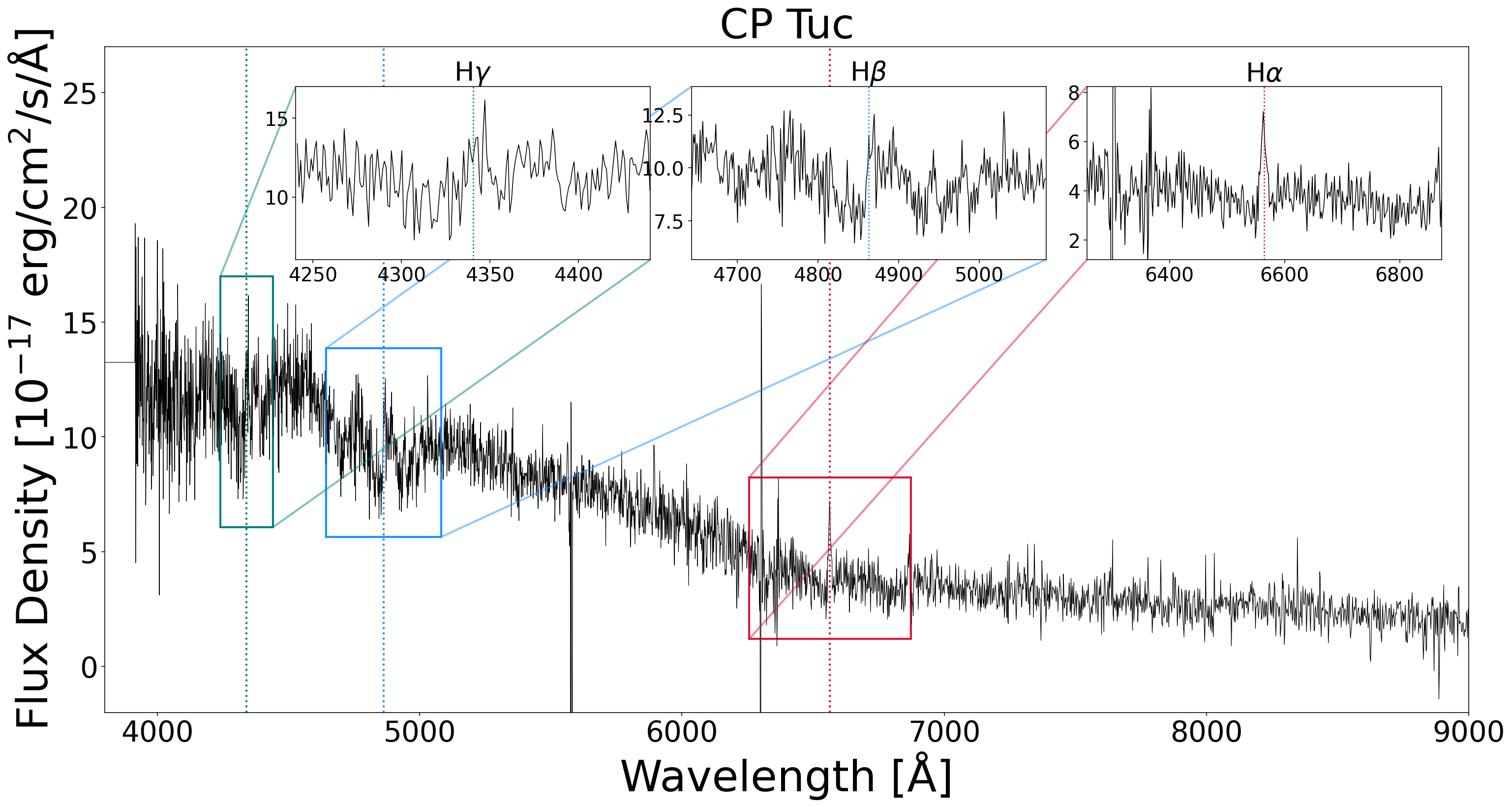}
    \includegraphics[width=0.49\textwidth]{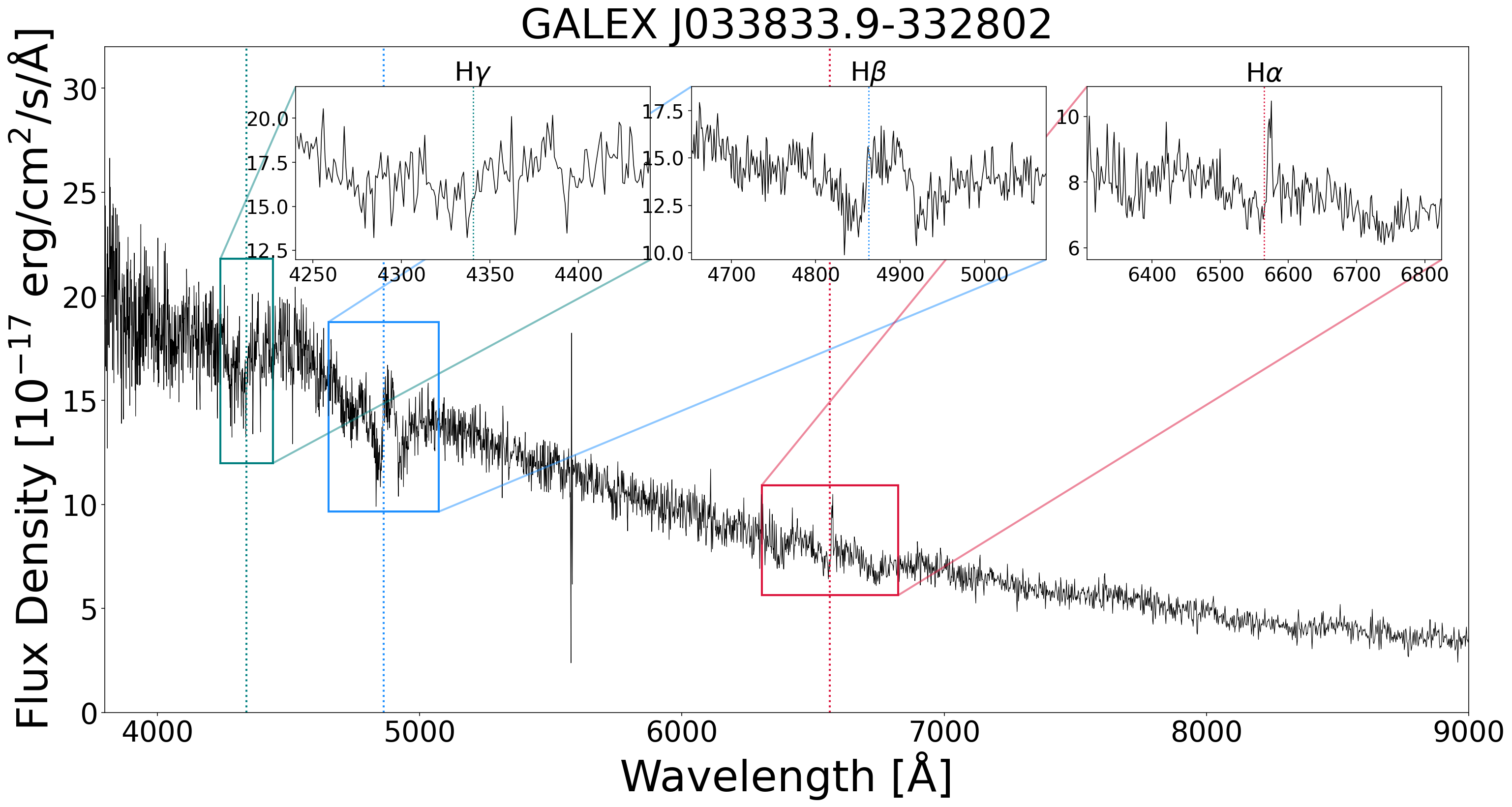}
    
    \vspace{0.4cm}
    
    \includegraphics[width=0.49\textwidth]{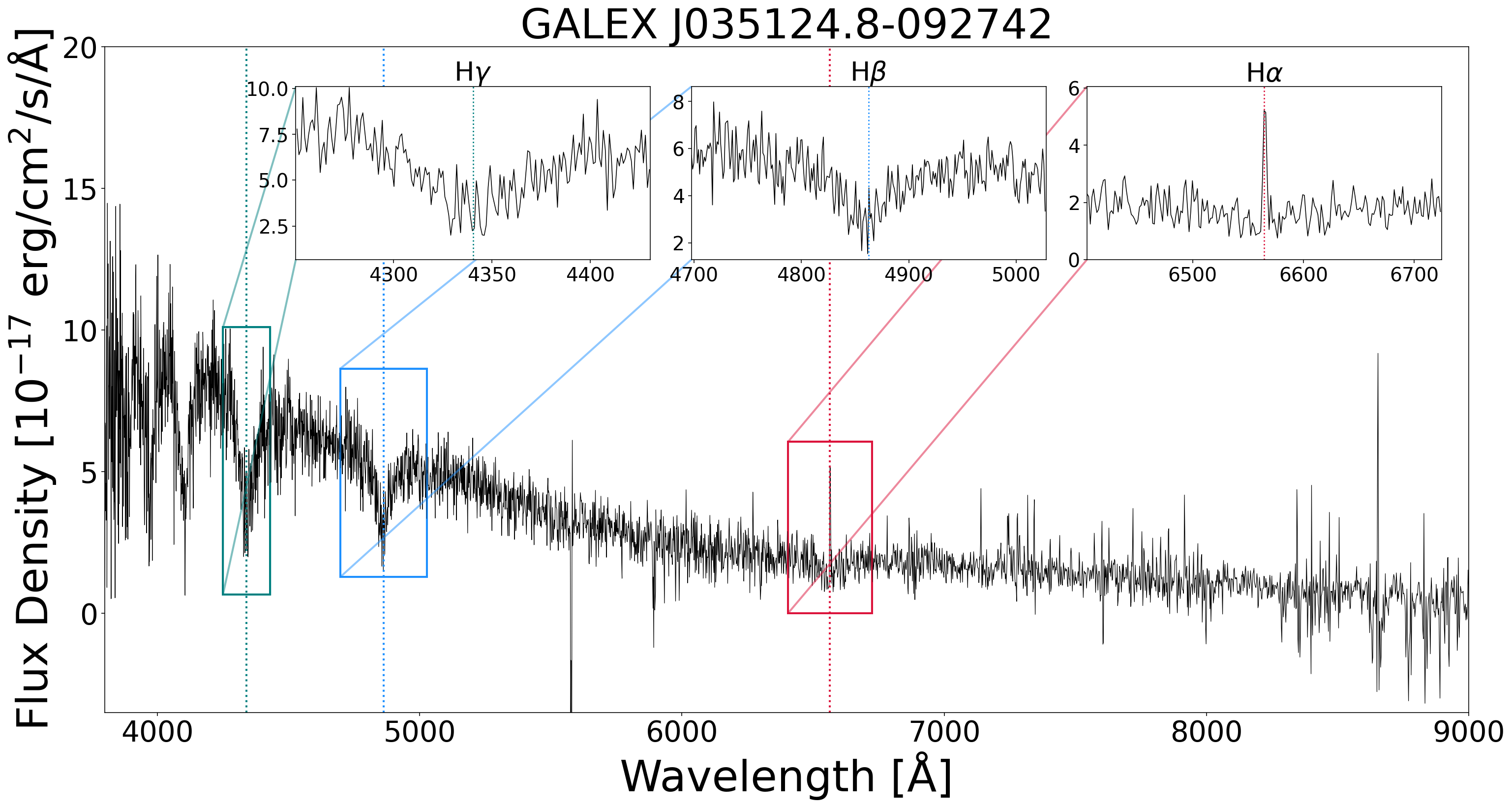}
    
    \caption{SDSS-V spectra of candidates for PBs approaching detachment. Each panel shows the full spectrum together with zoomed Balmer line regions.}
    
    \label{fig:detached_PBs_panel}
\end{figure}

\end{appendix}
\end{document}